\documentclass[reprint, pre, amssymb, amsmath, amsfonts, a4paper, aps, floatfix,longbibliography, superscriptaddress, showpacs]{revtex4-1} 

\usepackage{float}
\usepackage{amssymb, amsmath, amsfonts}
\usepackage{graphicx}
\usepackage[toc,page]{appendix}
\usepackage{units}
\usepackage{xcolor}

\usepackage{color}
\definecolor{Red}{rgb}{1,0,0}
\definecolor{blue}{rgb}{0,0,1}

\newcommand{\Epi}{\affiliation{Department of Epileptology, University of Bonn, Venusberg Campus 1, 53127~Bonn, Germany}}
\newcommand{\HISKP}{\affiliation{Helmholtz Institute for Radiation and Nuclear Physics, University of Bonn, Nussallee~14--16, 53115~Bonn, Germany}}
\newcommand{\IZKS}{\affiliation {Interdisciplinary Centre for Complex Systems, University of Bonn, Br\"uhler Stra\ss{}e~7, 53175~Bonn, Germany}}
\newcommand{\ICBM}{\affiliation {Department of Physics, Sharif University of Technology, 11365-9161, Tehran, Iran}}
\newcommand{\RNS}{\affiliation{Institute of Physics and ForWind, Carl von Ossietzky University of Oldenburg,\\ Carl-von-Ossietzky-Stra\ss{}e~9--11, 26111~Oldenburg, Germany}}
\newcommand{\THP}{\affiliation{Institute for Theoretical Physics, University of Cologne, 50937 K\"oln, Germany}}
\newcommand{\FZJ}{\affiliation{Forschungszentrum J\"ulich, Institute for Energy and Climate Research - Systems Analysis and Technology Evaluation (IEK-STE), 52428 J\"ulich, Germany}}

\newcommand{\nn}{\nonumber}

\newcommand*\diff{\mathop{}\!\mathrm{d}}
\renewcommand{\vec}[1]{\boldsymbol{#1}}

\def\be{\begin{equation}}
\def\ee{\end{equation}}
\def\bea{\begin{eqnarray}}
\def\eea{\end{eqnarray}}
\def\beqa{\begin{equation}\begin{array}{l}}
\def\eeqa{\end{array}\end{equation}}
\def\bsub{\begin{subequations}}
\def\esub{\end{subequations}}
\def\bce{\begin{center}}
\def\ece{\end{center}}

\newcommand{\dt}{\diff t}

\def\barr{\left(\begin{array}{c}}
\def\earr{\end{array}\right)}
\def\bmat{\left(\begin{array}{cc}}
\def\emat{\end{array}\right)}

\newcommand{\Mtwo}[2]{\mathcal{M}^{[#1, #2]}}
\newcommand{\ftwo}[2]{\ensuremath{f}^{[#1, #2]}}

\renewcommand{\d}{\mathrm{d}}

\begin{document}

\title{Analysis and data-driven reconstruction of bivariate jump-diffusion processes}

\author{Leonardo Rydin Gorj\~ao}
\email{\texttt{l.rydin.gorjao@fz-juelich.de}}
\Epi \HISKP \FZJ \THP

\author{Jan Heysel}
\email{\texttt{jan.heysel@uni-bonn.de}}
\Epi \HISKP

\author{Klaus Lehnertz}
\email{\texttt{klaus.lehnertz@ukbonn.de}}
\Epi \HISKP \IZKS

\author{M. Reza Rahimi Tabar}
\email{\texttt{tabar@uni-oldenburg.de}}
\RNS \ICBM

\begin{abstract}
We introduce the bivariate jump-diffusion process, comprising two-dimensional diffusion and two-dimensional jumps, that can be coupled to one another.
We present a data-driven, non-parametric estimation procedure of higher-order (up to 8) Kramers--Moyal coefficients that allows one to reconstruct
relevant aspects of the underlying jump-diffusion processes and to recover the underlying parameters.
The procedure is validated with numerically integrated data using synthetic bivariate time series from continuous and discontinuous processes.
We further evaluate the possibility of estimating the parameters of the jump-diffusion model via data-driven analyses of the higher-order Kramers--Moyal coefficients, and the limitations arising from the scarcity of points in the data or disproportionate parameters in the system.

\end{abstract}

\pacs{%
}
\maketitle


\section{Introduction}
Research over the last two decades has demonstrated the high suitability of the network paradigm in advancing our understanding of natural and man-made complex dynamical systems~\cite{boccaletti2006,arenas2008,bullmore2011,barthelemy2011,newman2012,holme2012,Kivela2014}.
With this paradigm, a system component is represented by a vertex and interactions between components are conveyed by edges connecting vertices, and graph theory provides a large repertoire of methods to characterize networks on various scales.

Characterizing properties of interactions using the knowledge of the dynamics of each of the components is key to understanding real-world systems.
To achieve this goal, a large number of time-series-analysis methods has been developed that originate from synchronization theory, nonlinear dynamics, information theory, and from statistical physics (for an overview, see Refs.~\cite{pikovsky2001,kantz2003,pereda2005,hlavackova2007,marwan2007,lehnertz2009b,lehnertz2011,stankovski2017}).
Some of these methods make rather strict assumptions about the dynamics of network components generating the time series and many approaches preferentially focus on the low-dimensional deterministic part of the dynamics.

Real-world systems, however, are typically influenced by random forcing and interactions between constituents are highly non-linear, which results in very complex, stochastic, and non-stationary system behavior that exhibits both deterministic and stochastic features.
Aiming at determining characteristics and strength of fluctuating forces as well as at assessing properties of non-linear interactions, the analysis of such systems is associated with the problem of retrieving a stochastic dynamical system from measured time series.
There is a substantial existing literature~\cite{Risken1996,Kampen1981,Friedrich2011,Tabar-book2019} for the modeling of complex dynamical systems which employs the conventional Langevin equation that is based on the first- and second-order Kramers--Moyal (KM) coefficients, known as drift and diffusion terms.
All functions and parameters of this modeling can be found directly from the measured time series employing a widely used non-parametric approach.
There are by now only few studies that make use of this ansatz to characterize interactions between stochastic processes~\cite{Prusseit2008a,lehle2013stochastic,scholz2017parameter,wahl2016granger,lind2010extracting}.

Despite its successful application in diverse scientific fields, growing evidence indicates that the continuous stochastic modeling of time series of complex systems (white-noise-driven Langevin equation) should account for the presence of discontinuous jump components~\cite{weissman1988,Bakshi1997,Duffie2000,andersen2002,Das2002,Johannes2004,Cai2009,anvari2016,lehnertz2018,Tabar-book2019}.
In this context, the jump-diffusion model~\cite{chudley1961,merton1976,hall1981,Giraudo1997} was shown to provide a theoretical tool to study processes of known and unknown nature that exhibit jumps.
It allows one to separate the deterministic drift term as well as different stochastic behaviors, namely diffusive and jumpy behavior~\cite{anvari2016,lehnertz2018,Tabar-book2019}.
Moreover, all of the unknown functions and coefficients of a dynamical stochastic equation that describe a jump-diffusion process can be derived directly from measured time series.
This approach involves estimating higher-order ($\geq 3$) KM coefficients and it provides an intuitive physical meaning of these coefficients.

The focus of this paper is to introduce a method to investigate bivariate time-series with discontinuous jump components.
We begin with an overview of bivariate diffusion processes that exhibit the known relation between the parameters and functions in stochastic modeling and the KM coefficients.
Exemplary processes are portrayed, and we propose a measure to judge the quality of our reconstruction procedure.
We then present bivariate jump-diffusion processes alongside with the associated KM expansion~\cite{anvari2016}, and investigate the suitability of our reconstruction procedure using various examples.
We conclude this paper by summarizing our findings.

\section{State of the art: Bivariate jump-diffusion model}
A bivariate jump-diffusion process can be modeled via \cite{anvari2016,Tabar-book2019}
\begin{equation}\label{eq:model}
\begin{aligned}
    \overbrace{\begin{pmatrix}
    \mathrm{d}y_1(t) \\ \mathrm{d}y_2(t)
    \end{pmatrix}}^{\displaystyle\vec{y}}
&=\underbrace{
    \overbrace{\begin{pmatrix}
    N_1 \\ N_2
    \end{pmatrix}}^{\displaystyle\vec{N}}}_{\text{drift}}
    \d t
+ \underbrace{
    \overbrace{\begin{pmatrix}
    g_{1,1} & g_{1,2} \\
    g_{2,1} & g_{2,2}
    \end{pmatrix}}^{\displaystyle\vec{g}}
    \overbrace{\begin{pmatrix}
    \d w_1 \\ \mathrm{d}w_2
    \end{pmatrix}}^{\displaystyle\vec{\d w}}
    }_{\text{diffusion}}
+ \\
~&~\qquad\qquad\qquad\qquad\underbrace{
    \overbrace{\begin{pmatrix}
    \xi_{1,1} & \xi_{1,2} \\
    \xi_{2,1} & \xi_{2,2}
    \end{pmatrix}}^{\displaystyle\vec{\xi}}
    \overbrace{\begin{pmatrix}
    \mathrm{d}J_1 \\ \mathrm{d}J_2
    \end{pmatrix}}^{\displaystyle\vec{\d J}}
    }_{\text{Poissonian jumps}},
\end{aligned}
\end{equation}
where all the elements of vectors $\vec{N}$, $\vec{\d J}$, and $\vec{\d w}$ as well as of matrices $\vec{g}$ and $\vec{\xi}$ may, in general, be state- and time-dependent (dependencies not shown for convenience of notation).
The drift coefficient is a two-dimensional vector $\vec{N} = (N_1, N_2)$ with $\vec{N} \in \mathbb{R}^2$, where each dimension of $\vec{N}$, i.e., $N_i$, may depend on $y_1(t)$ and $y_2(t)$.
The diffusion coefficient takes a matrix $\vec{g} \in \mathbb{R}^{2\times 2}$.
The two Wiener processes $\vec{w}=(w_1,w_2)$ act as independent Brownian noises for the state variables $y_1(t)$ and $y_2(t)$.
The diagonal elements of $\vec{g}$ comprise the diffusion coefficients of self-contained stochastic diffusive processes, and the off-diagonal elements represent interdependencies between the two Wiener processes, i.e., they result from an interaction between the two processes.
Each single-dimensional stochastic process element $\d w_i$ of $\vec{\d w}$ is an increment of a Wiener process, with $\langle \d w_i \rangle = 0, \langle \d w_i^2 \rangle = \d t, \forall i$.
The discontinuous jump terms are contained in $\vec{\xi} \in \mathbb{R}^{2\times 2}$ and $\vec{\d J} \in \mathbb{N}^{2}$, where $\vec{\d J}$ represents a two-dimensional Poisson process.
These are Poisson-distributed jumps with an average jump rate $\vec{\lambda} \in \mathbb{R}^2$ in unit time $t$.
The average expected number of jumps of each jump process  $J_i$ in a timespan $t$ is $\lambda_i t$.
The jump amplitudes $\vec{\xi}$ are Gaussian distributed with zero mean and standard deviation $\xi_{i,j}$.

We here consider merely autonomous systems.
Non-ergodic problems are beyond the scope of this work, and a more delicate approach to both bivariate stochastic processes would be needed.

\section{Bivariate diffusion processes}
Let us begin with bivariate diffusion processes, for which the model takes the form
\begin{equation}\label{eq:model_reduced}
\begin{aligned}
    \overbrace{\begin{pmatrix}
    \mathrm{d}y_1(t) \\ \mathrm{d}y_2(t)
    \end{pmatrix}}^{\displaystyle\vec{y}}
 =\underbrace{
    \overbrace{\begin{pmatrix}
    N_1 \\ N_2
    \end{pmatrix}}^{\displaystyle\vec{N}}
    \d t}_{\text{drift}}
+ \underbrace{
    \overbrace{\begin{pmatrix}
    g_{1,1} & g_{1,2} \\
    g_{2,1} & g_{2,2}
    \end{pmatrix}}^{\displaystyle\vec{g}}
    \overbrace{\begin{pmatrix}
    \d w_1 \\ \mathrm{d}w_2
    \end{pmatrix}}^{\displaystyle\vec{\d w}}
    }_{\text{diffusion}}
\end{aligned}.
\end{equation}
The model consists of six functions, two for the drift coefficients and four for the diffusion coefficients.
Given a bivariate diffusion process, can we reconstruct the aforementioned parameters strictly from data?
Extensive work exists on this matter~\cite{Friedrich2011}, especially covering purely diffusion processes, and we will use these now as a stepping stone to jump-diffusion processes.
Understanding the working and contingencies of reconstructing the parameters of a diffusion process (Eq.~\eqref{eq:model_reduced}) will serve as a gateway to understand how a similar procedure awards equivalent measures for jump-diffusion processes.
We address the aforementioned question firstly by revisiting the mathematical foundation that allows one to recover, strictly from data, the drift $\vec{N}$ and diffusion $\vec{g}$ coefficients.
Subsequently, we numerically integrate diffusion processes with a priori fixed values of the drift $\vec{N}$ and diffusion $\vec{g}$ coefficients and aim at retrieving these values strictly from the generated data (Euler--Mayurama scheme with a time sampling of $10^{-3}$ over a total of $10^5$ time units, i.e., $10^8$ number of data points).
If the actual and retrieve values match, the reconstruction method is effective.

A stochastic process has a probabilistic description given by the master equation~\cite{Risken1996,Tabar-book2019}.
It does not describe a specific stochastic processes in itself, but the probabilistic evolution of the process in time.
The master equation accepts an expansion in terms, the KM expansion, that allows for a purely differential description of the process.
More importantly, the coefficients of the expansion, known as KM coefficients, entail directly a relation to the aforementioned parameters of a stochastic process given by Eq.~\eqref{eq:model}.
The exact relation will be given below.
\begin{figure*}[t]
\includegraphics[width=.48\linewidth]{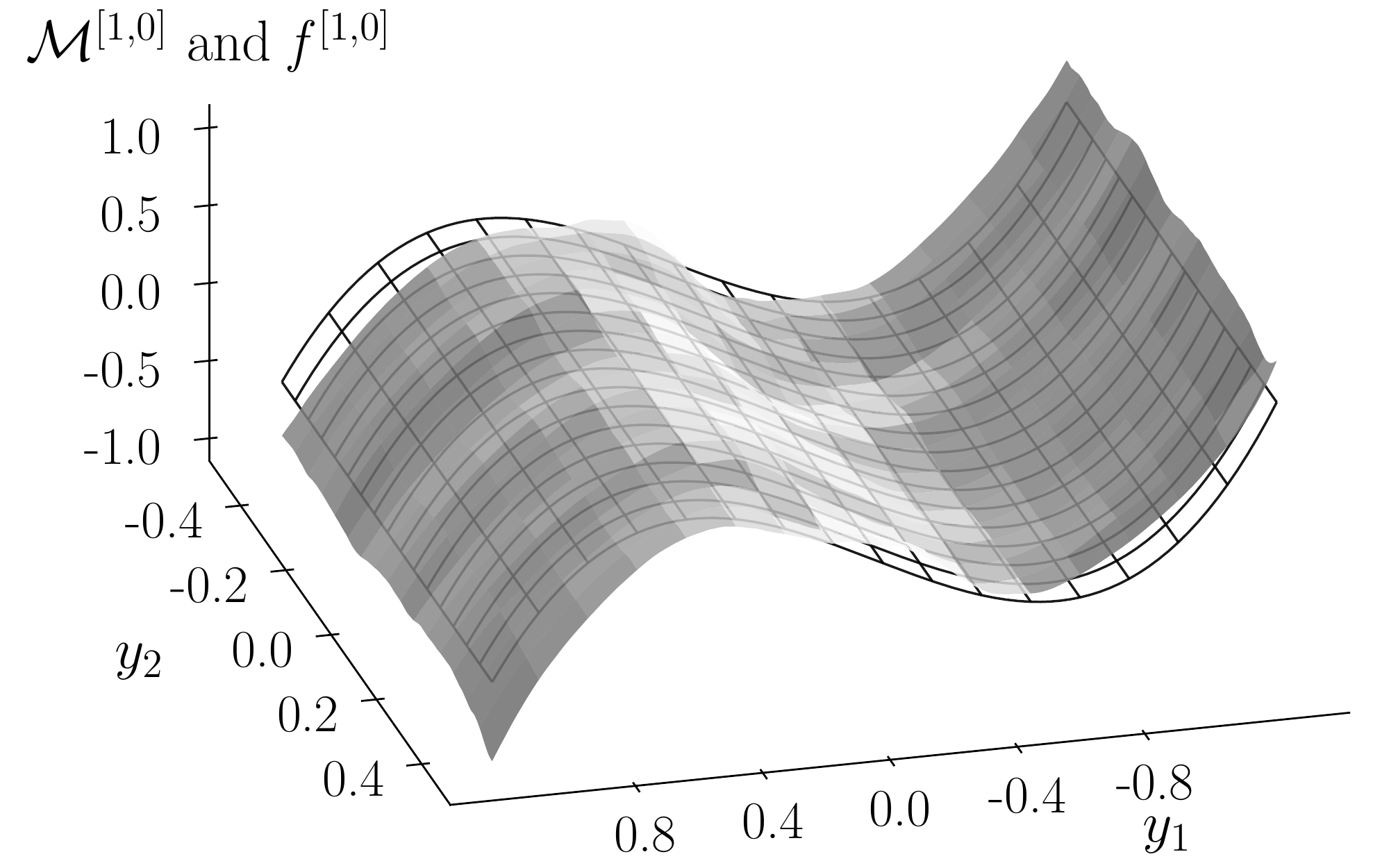}
\includegraphics[width=.48\linewidth]{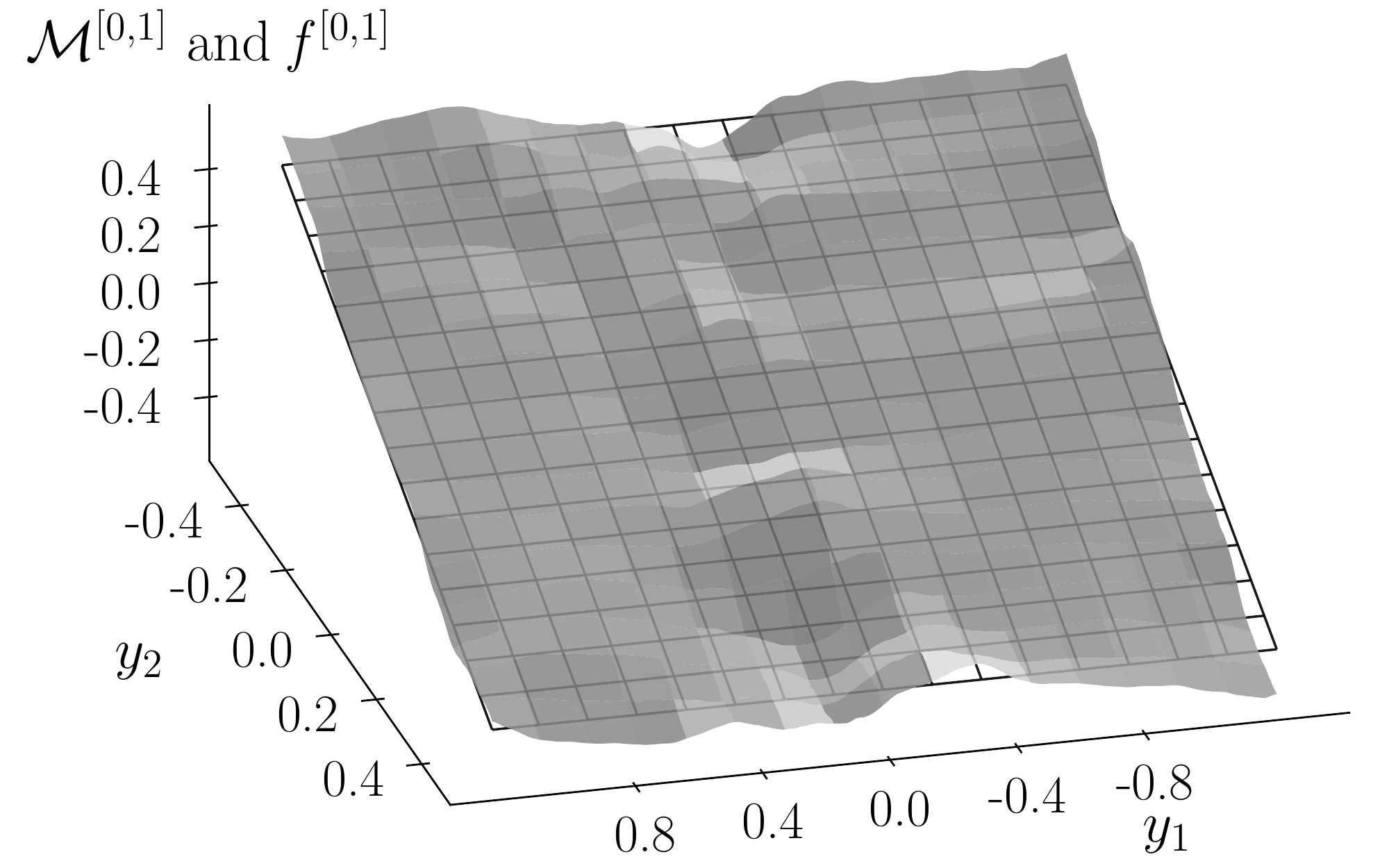}
\includegraphics[width=.48\linewidth]{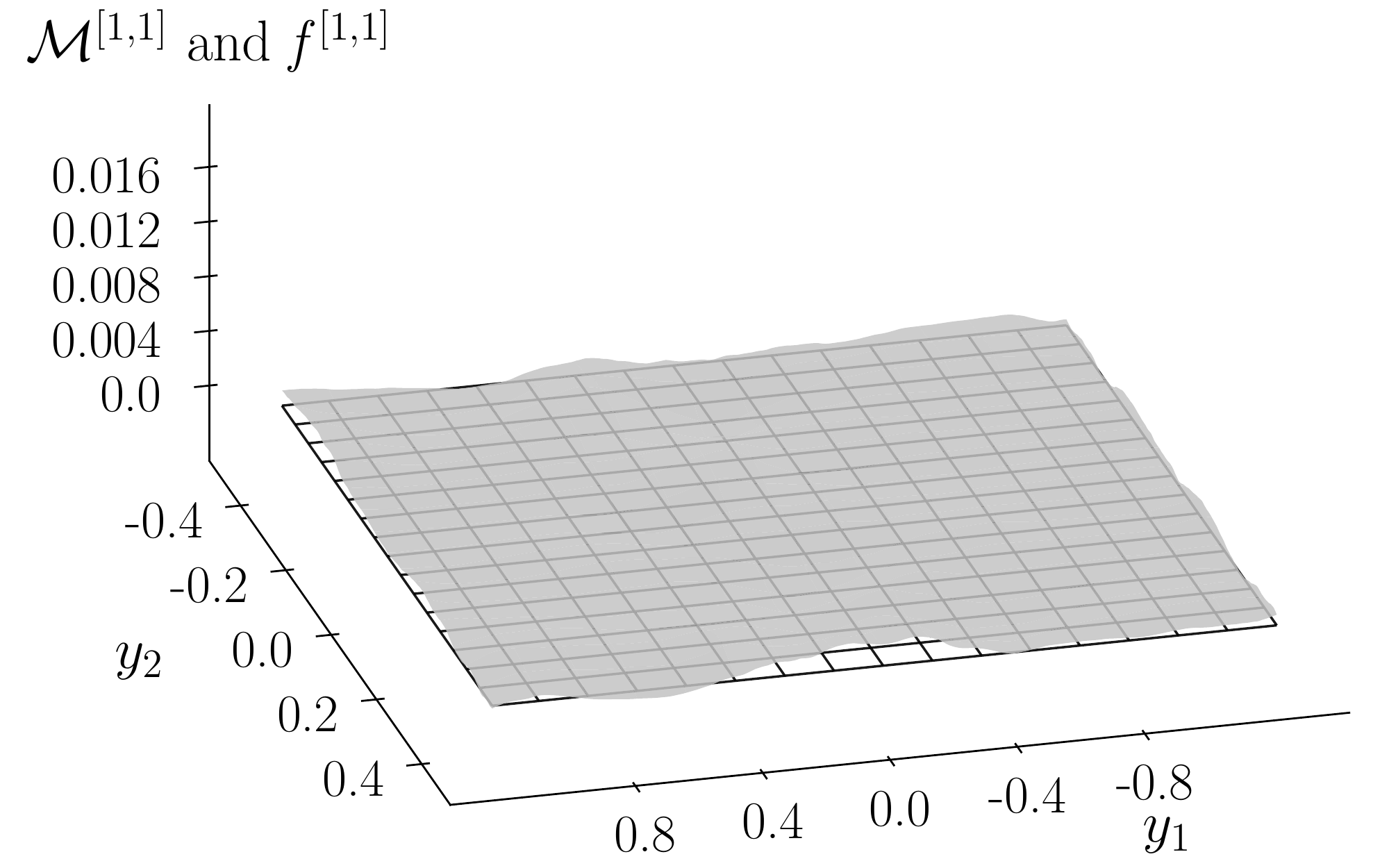}
\includegraphics[width=.48\linewidth]{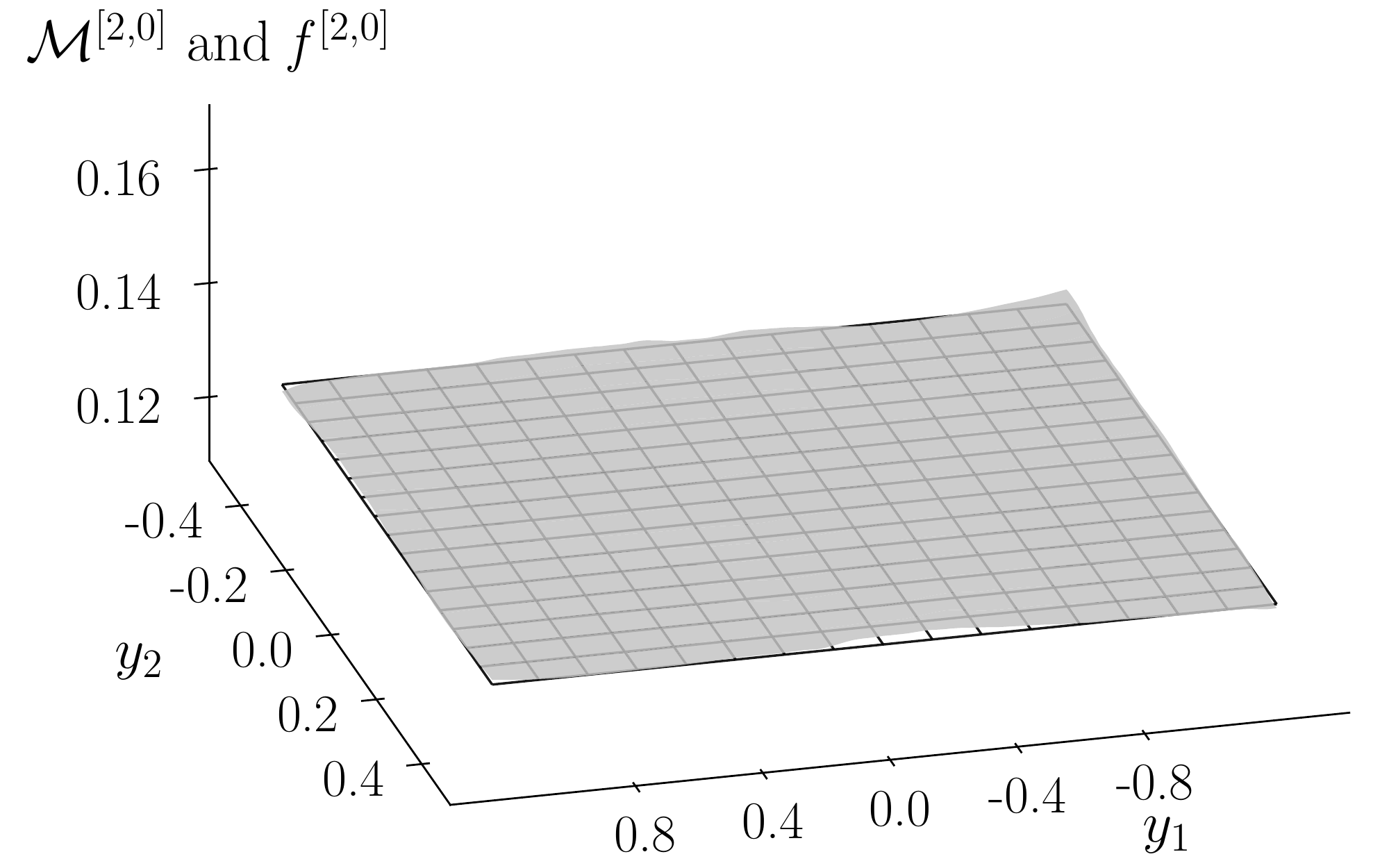}
\caption{Two-dimensional Kramers--Moyal coefficients $\Mtwo{\ell}{m}$ for two independent diffusion processes given by Eq.~\eqref{eq:example_cubic_constant_noise}.
The uncovered KM surfaces match the expectation, the cubic term in the drift term $N_1 = -x_1^3+x_1$ along the first variable is visible on the top-left panel, $\Mtwo{1}{0}$, and the negative-slope surface in the top-right panel, $\Mtwo{0}{1}$.
The flat surfaces reproduce as well the expected form of the constant terms involved in the diffusion terms for $\Mtwo{2}{0}$ on the bottom-right.
Moreover, the bottom-left panel $\Mtwo{1}{1}$, which accounts for the couplings terms of all diffusion terms, is also zero almost everywhere, as expected, given $g_{1,2}$ and $g_{2,1}$ are zero.
In each panel, the theoretical expected surface, given by Eqs.~\eqref{eq:drift_km2_diff} and~\eqref{eq:diffusion_km2_diff}, is indicated by a grid, with $\ftwo{\ell}{m}$ denoting the respective theoretical values introduced in the model.}
\label{fig:Cubic_constant_noise}
\end{figure*}
There is although an important detail in what regards the KM coefficients: they are in themselves not constants, but functions on the underlying space, or in other words, a scalar field, and for our purposes here, they can be understood as two-dimensional surfaces. We will denote these as KM surfaces.

Lastly, and more familiar, the Fokker--Planck equation is a truncation of the KM expansion at second order.
It is especially relevant given its connection to physical processes and the Pawula theorem~\cite{Pawula1967}.
The Pawula theorem ensures that the truncation is not ill-suited for the underlying process if the fourth-order KM coefficient approaches zero in the limit $\d t \to 0$.
It is now crucial to understand that the theorem holds for a one-dimensional process, and we are not aware of a proof for higher dimensions.
This contrasts the common notion that studying only the first two KM coefficients of two- or higher-dimensional processes is sufficient (see Refs.~\cite{Prusseit2007,anvari2016,lehnertz2018} and references therein).

The KM coefficients $\Mtwo{\ell}{m}(x_1,x_2)\in\mathbb{R}^2$ of orders $(\ell, m)$ are defined as:
\begin{equation}
\begin{aligned}
&\Mtwo{\ell}{m}(x_1,x_2)=\\
&\lim_{\Delta t\to 0}\!\frac{1}{\Delta t}
\!\!\int \!\!(y_1(t\!+\!\Delta t)\!-\!y_1(t))^\ell(y_2(t\!+\!\Delta t)\!-y_2(t))^m \cdot \\
& \qquad P(y_1,y_2; t\!+\!\Delta t|y_1,y_2 ; t)|_{y_1(t)=x_1, y_2(t)=x_2}\d y_1\d y_2\nn,
\end{aligned}
\end{equation}
and can be obtained from bivariate time series $(y_1(t),y_2(t))$.
Theoretically, $\Delta t$ should take the limiting case of $\Delta t \to 0$, but the restriction of any measuring or storing devices---or the nature of the observables themselves---permits only time-sampled or discrete recordings.
The relevance and importance of adequate time-sampling was extensively studied and discussed in Ref.~\cite{lehnertz2018,Tabar-book2019}.

In the limiting case where $\Delta t$ is equivalent to the sampling rate of the data, the KM coefficients take the form
\begin{align}\label{eq:2D_KMC}
\Mtwo{\ell}{m}(x_1, x_2) &= \frac{1}{\Delta t} \langle \Delta y_1^{\ell} \Delta y_2^{m} |_{y_1(t)=x_1, y_2(t)=x_2}\rangle, \\
\Delta y_i &=  y_i(t+ \Delta t) - y_i(t).\nn
\end{align}

\begin{figure*}[t]
\includegraphics[width=.48\linewidth]{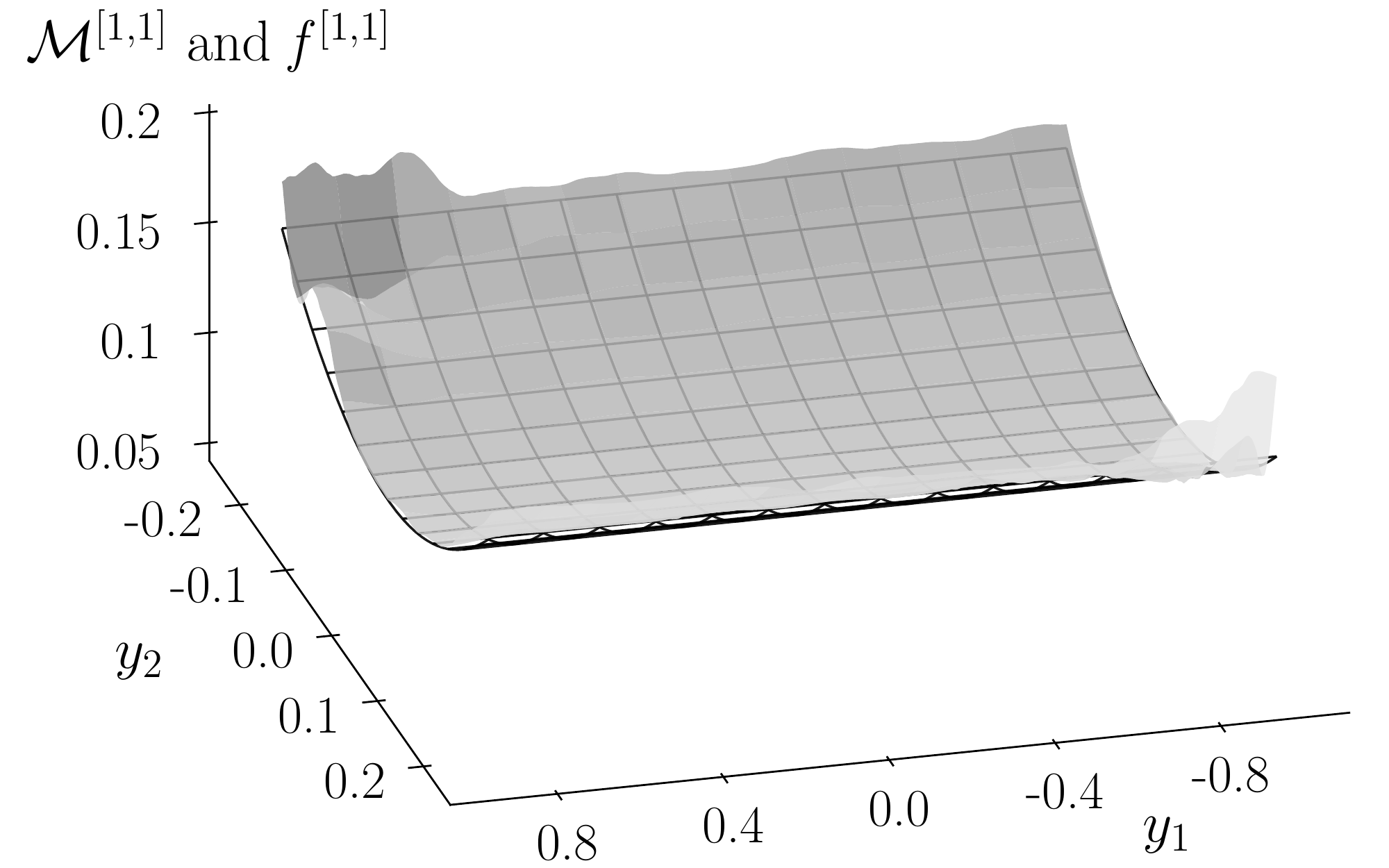}
\includegraphics[width=.48\linewidth]{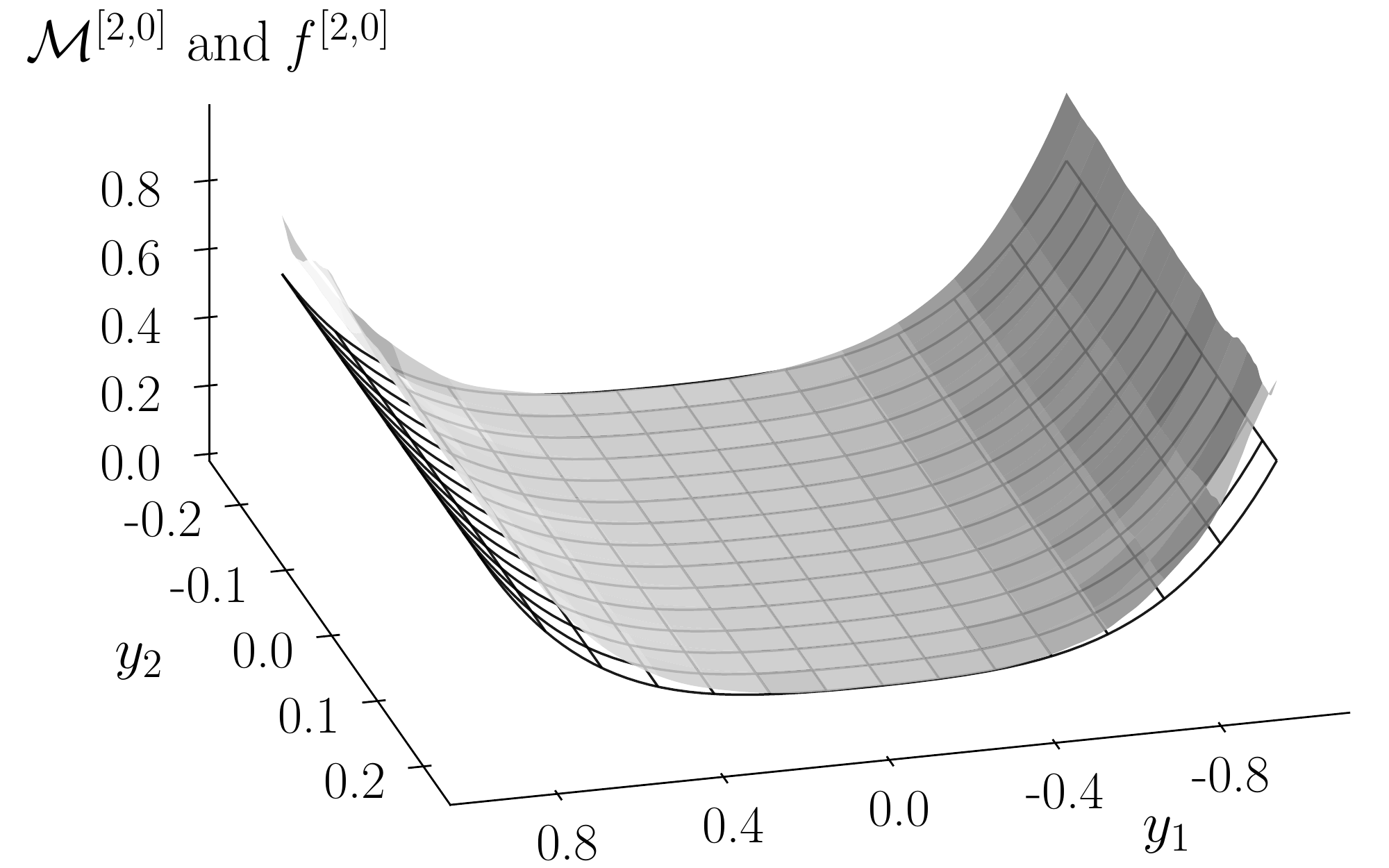}
\includegraphics[width=.48\linewidth]{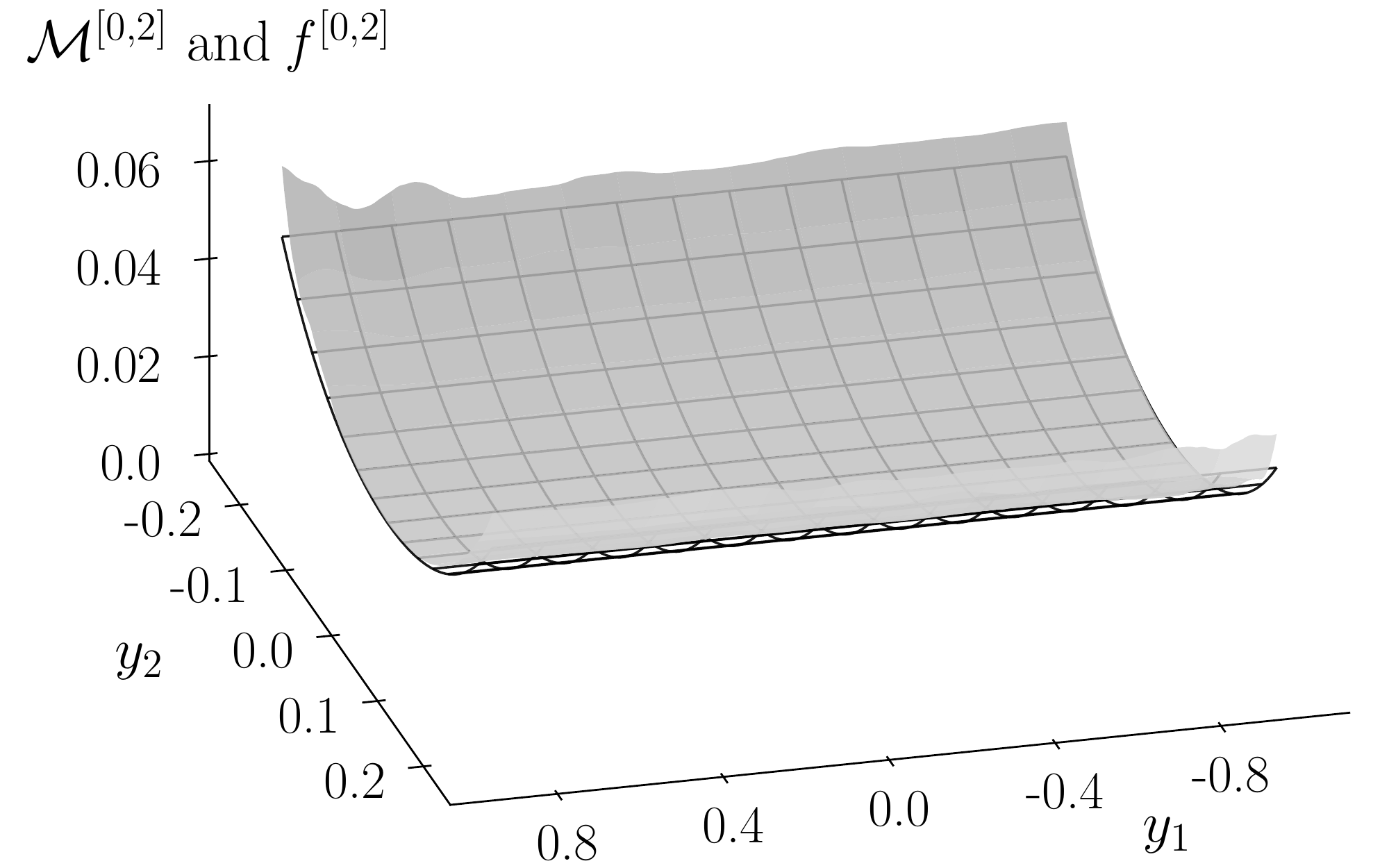}
\includegraphics[width=.48\linewidth]{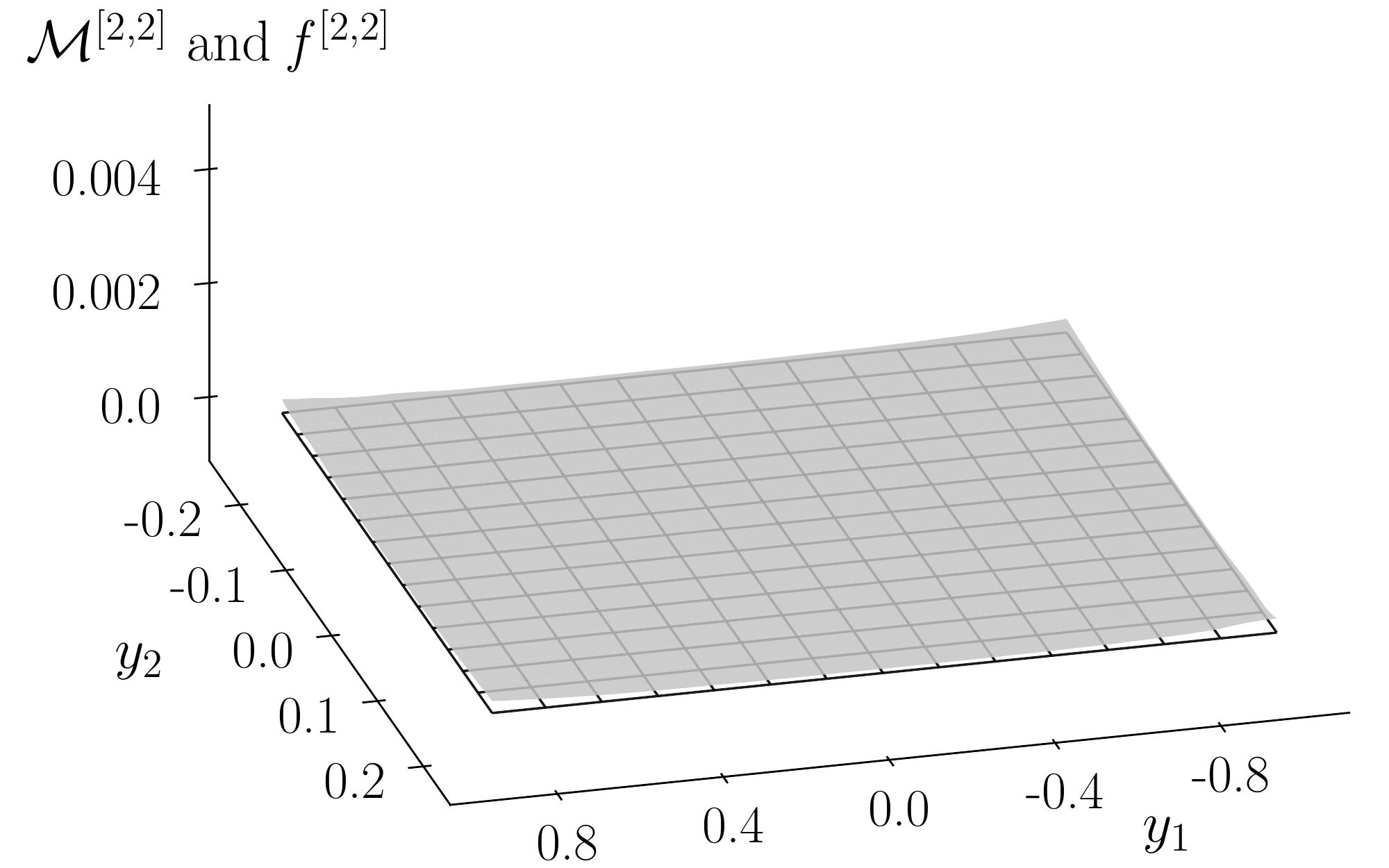}
\caption{Two-dimensional Kramers--Moyal (KM) coefficients $\Mtwo{\ell}{m}$ for two independent diffusion processes given by Eq.~\eqref{eq:example_cubic_weird_noise} and the theoretical expected functions $\ftwo{\ell}{m}$ associated with each coefficient, according to Eq.~\eqref{eq:diffusion_km2_diff}.
KM coefficients $\Mtwo{1}{1}$, $\Mtwo{2}{0}$, $\Mtwo{0}{2}$, and $\Mtwo{2}{2}$ exhibit the quadratic multiplicative dependencies of the diffusion terms.
In addition, $\Mtwo{1}{1}$ displays both an offset from zero as well as a quadratic shape, entailing the desired results emerging from Eq.~\eqref{eq:diffusion_km2_diff}, i.e., the noise-coupling term $g_{1,2}$ with $g_{2,2}$.
$\Mtwo{2}{0}$ displays an offset and has a minimum close to $g_{1,1}^2 / 2 + g_{1,2}^2 / 2 = 0.13$.
We also show the higher-order coefficient $\Mtwo{2}{2}$ and the corresponding theoretically expected value (given by Eq.~\eqref{eq:diffusion_km2_diff}), both of which vanish.
All obtained KM surfaces fit considerably well their theoretically expected ones ($V_\text{err}^{[1,1]}= 0.03$, $V_\text{err}^{[2,0]}=0.94$, $V_\text{err}^{[0,2]}=0.03$, $V_\text{err}^{[2,2]} < 0.01$; error volumes estimated over the displayed domain).}
\label{fig:Cubic_weird_noise}
\end{figure*}

The algebraic relations between the KM coefficients and functions in Eq.~\eqref{eq:model_reduced} are given by~\cite{anvari2016,Tabar-book2019}
\begin{align}
&\begin{aligned}\label{eq:drift_km2_diff}
\Mtwo{1}{0} &=\, N_1,\\
\Mtwo{0}{1} &=\, N_2,
\end{aligned}\\[.5em]
&\begin{aligned}\label{eq:diffusion_km2_diff}
\Mtwo{1}{1} &= \,g_{1,1} g_{2,1} ~+~ g_{1,2} g_{2,2}, \\
\Mtwo{2}{0} &=\, \left[\,g_{1,1}^2 ~+~ g_{1,2}^2  \right],\\
\Mtwo{0}{2} &=\, \left[g_{2,1}^2 ~+~ g_{2,2}^2\right].
\end{aligned}
\end{align}
An explicit derivation can be found in Appendix~\ref{App:1}.
Evidently, this under-determined set of five equations is insufficient to uncover the six functions of a general stochastic diffusion process.
One must bare this in mind, for the same issue will arise again when reconstructing jump-diffusion processes from data.
Nonetheless, under certain assumptions it is possible to reduced the dimension of the problem and therefore obtain a system of equation which is not under-determined.
Two methods for these cases are present in Ref.~\cite{Tabar-book2019} and another criterion will be presented later.

In order to relate the results obtained from studying the KM coefficients against the theoretical functions, we propose a method to assess the difference between the values of the theoretically expected functions and the estimated values of the KM coefficients.
Since for bivariate processes the KM coefficients are two-dimensional---as are the parameters of Eq.~\eqref{eq:model}---an adequate ``distance'' measure between the resulting two-dimensional surfaces is required.

Following Ref.~\cite{Prusseit2008a}, we propose a distance measure that allows for the variability of the density of data in some regions of the underlying space to be taken into consideration.
\begin{figure*}[t]
\includegraphics[width=.32\linewidth]{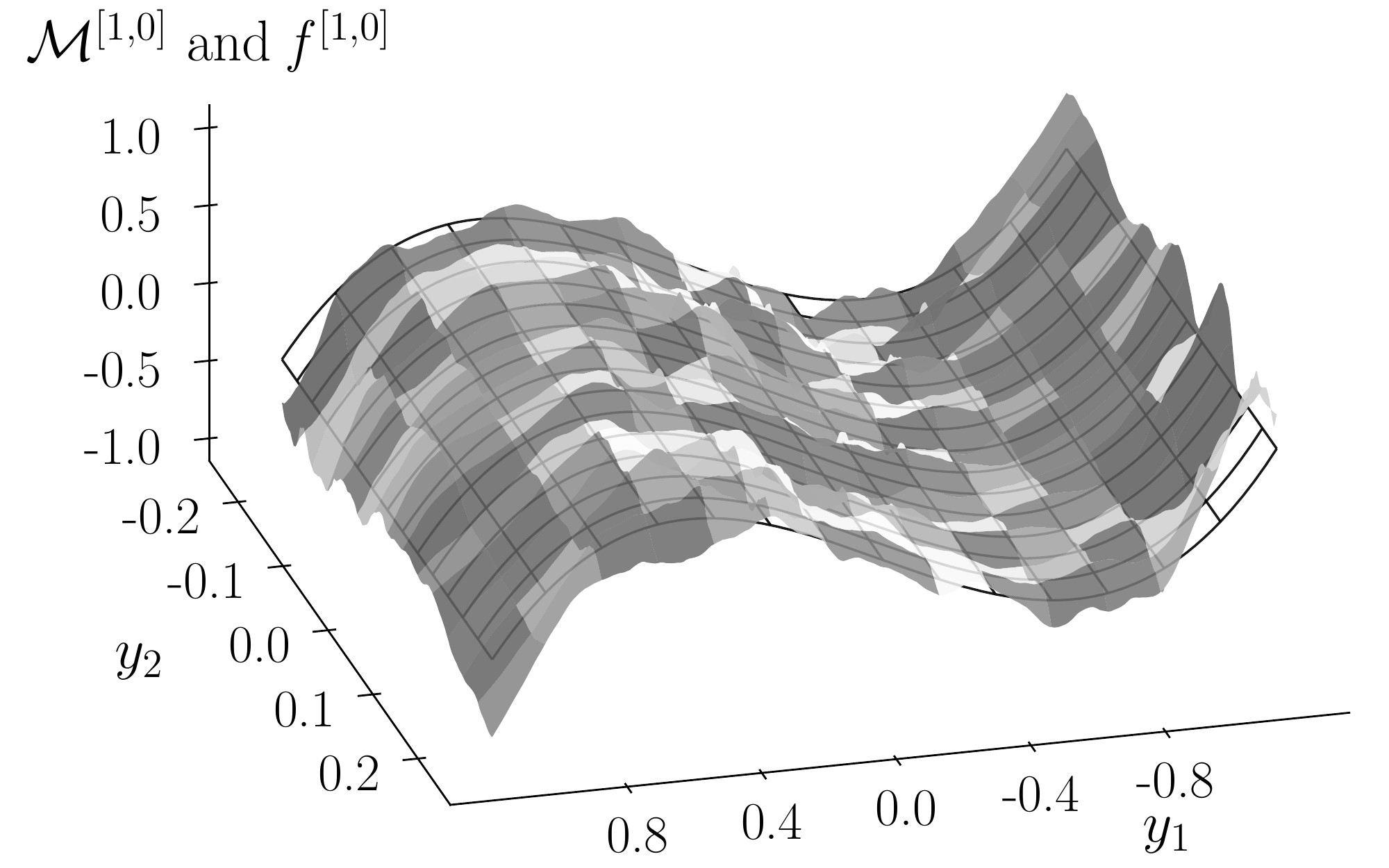}
\includegraphics[width=.32\linewidth]{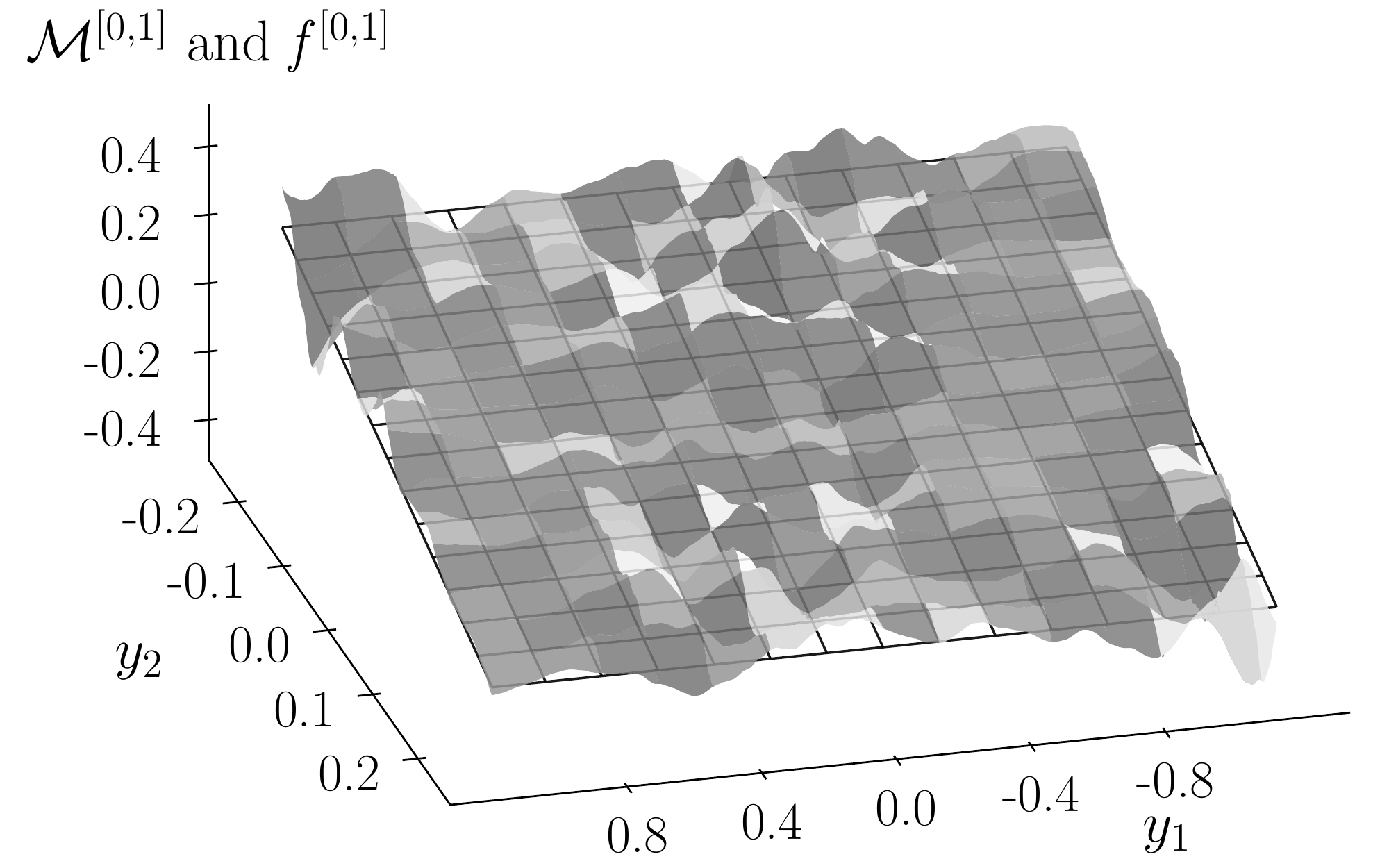}
\includegraphics[width=.32\linewidth]{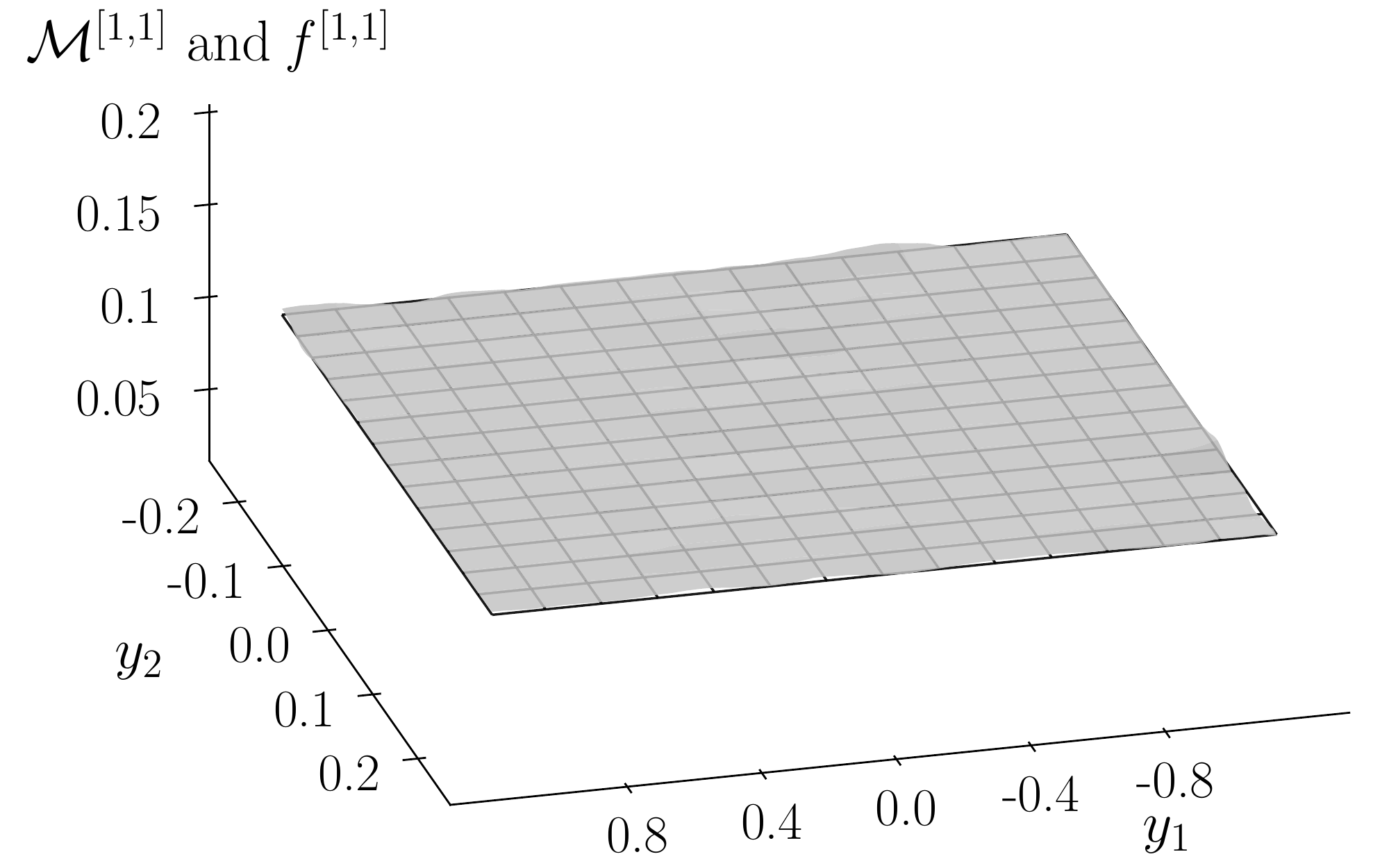}
\includegraphics[width=.32\linewidth]{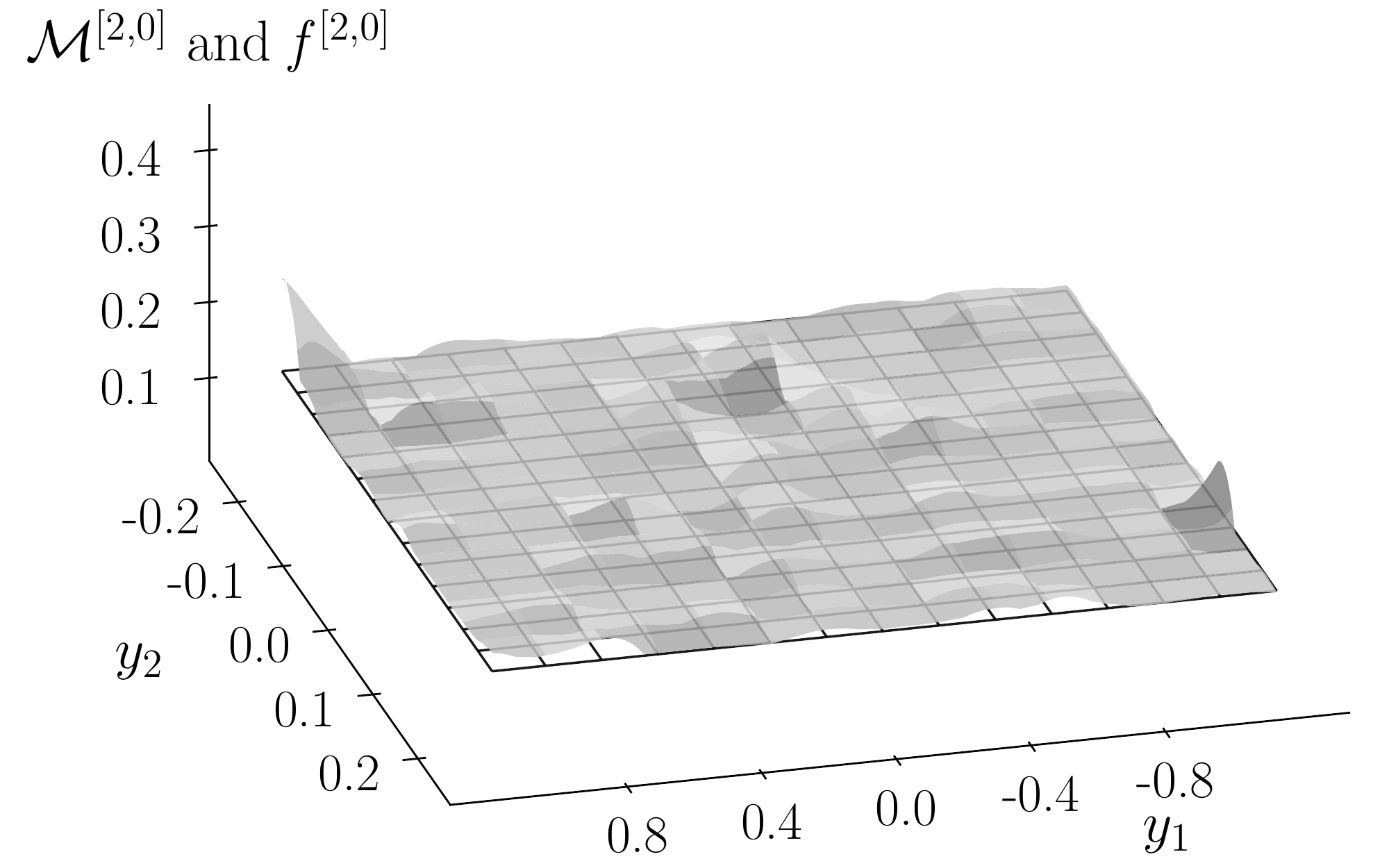}
\includegraphics[width=.32\linewidth]{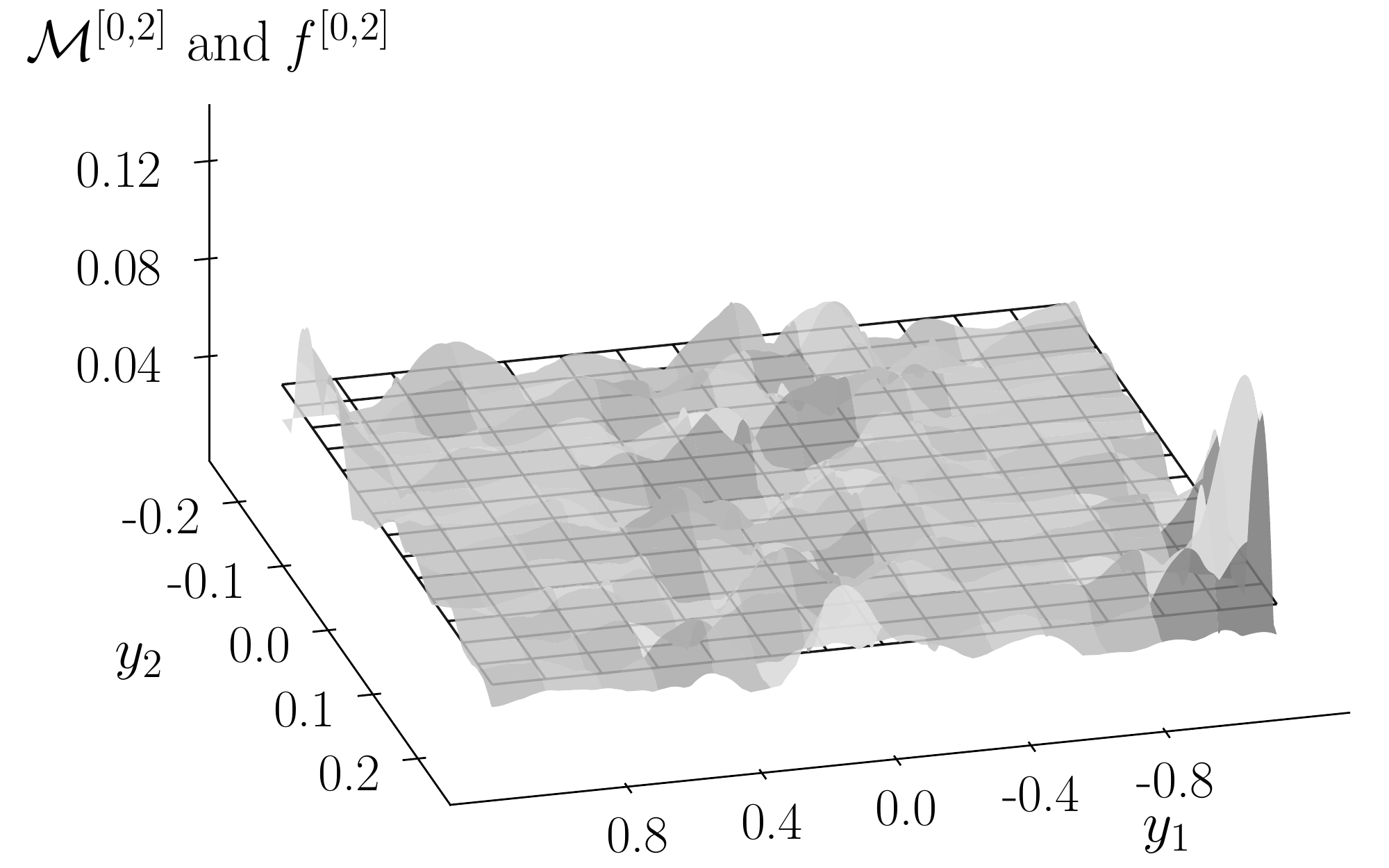}
\includegraphics[width=.32\linewidth]{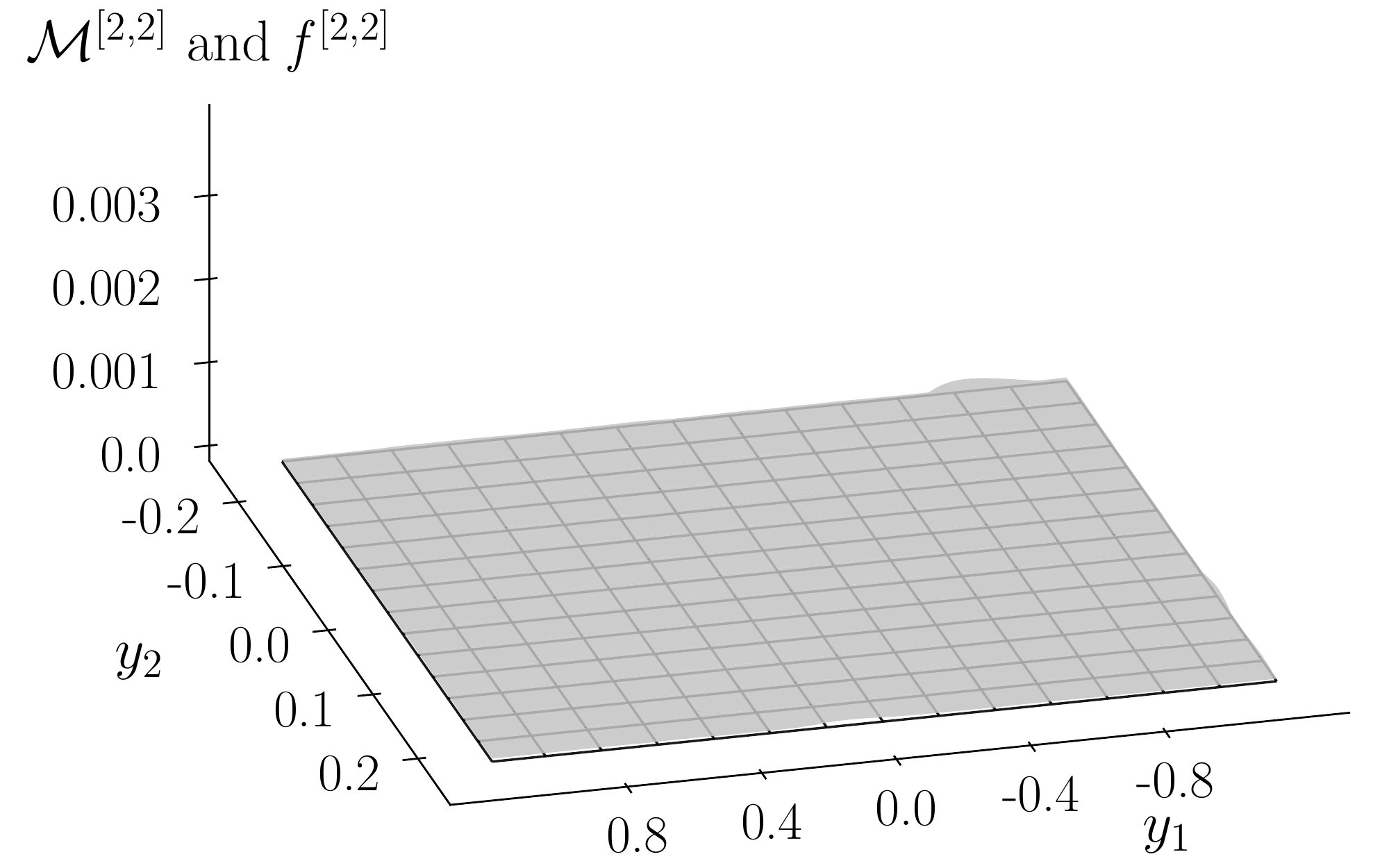}
\includegraphics[width=.32\linewidth]{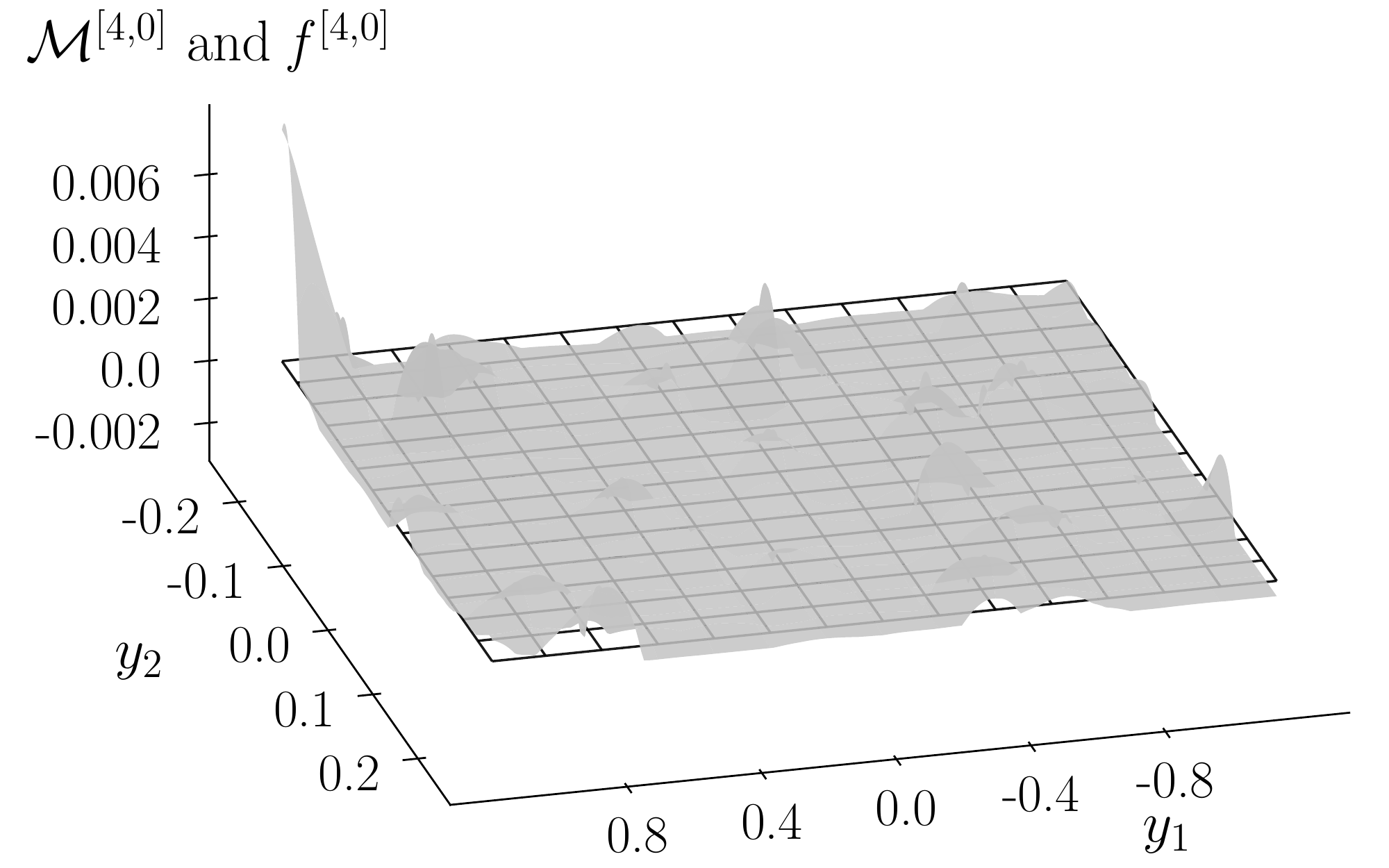}
\includegraphics[width=.32\linewidth]{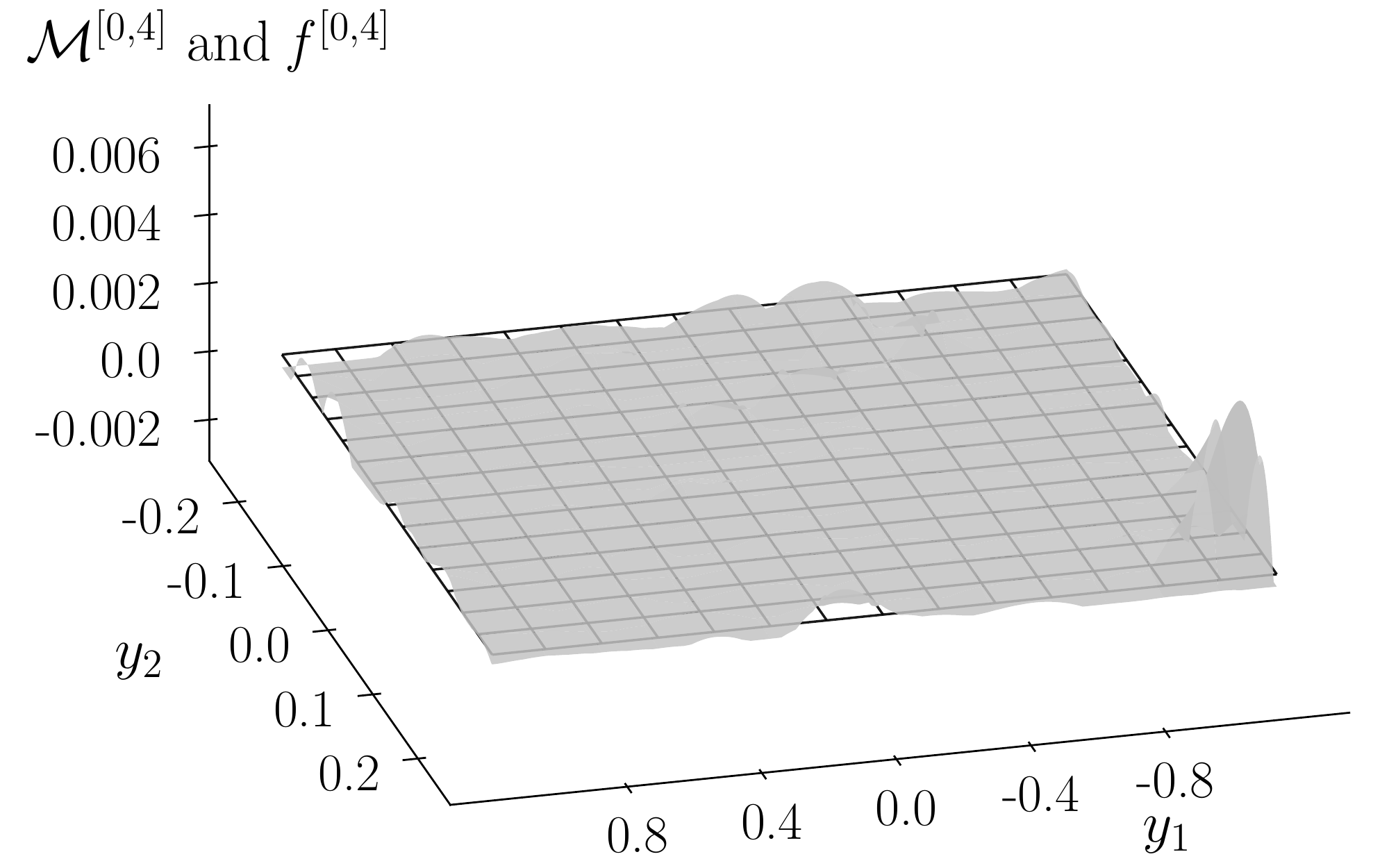}
\includegraphics[width=.32\linewidth]{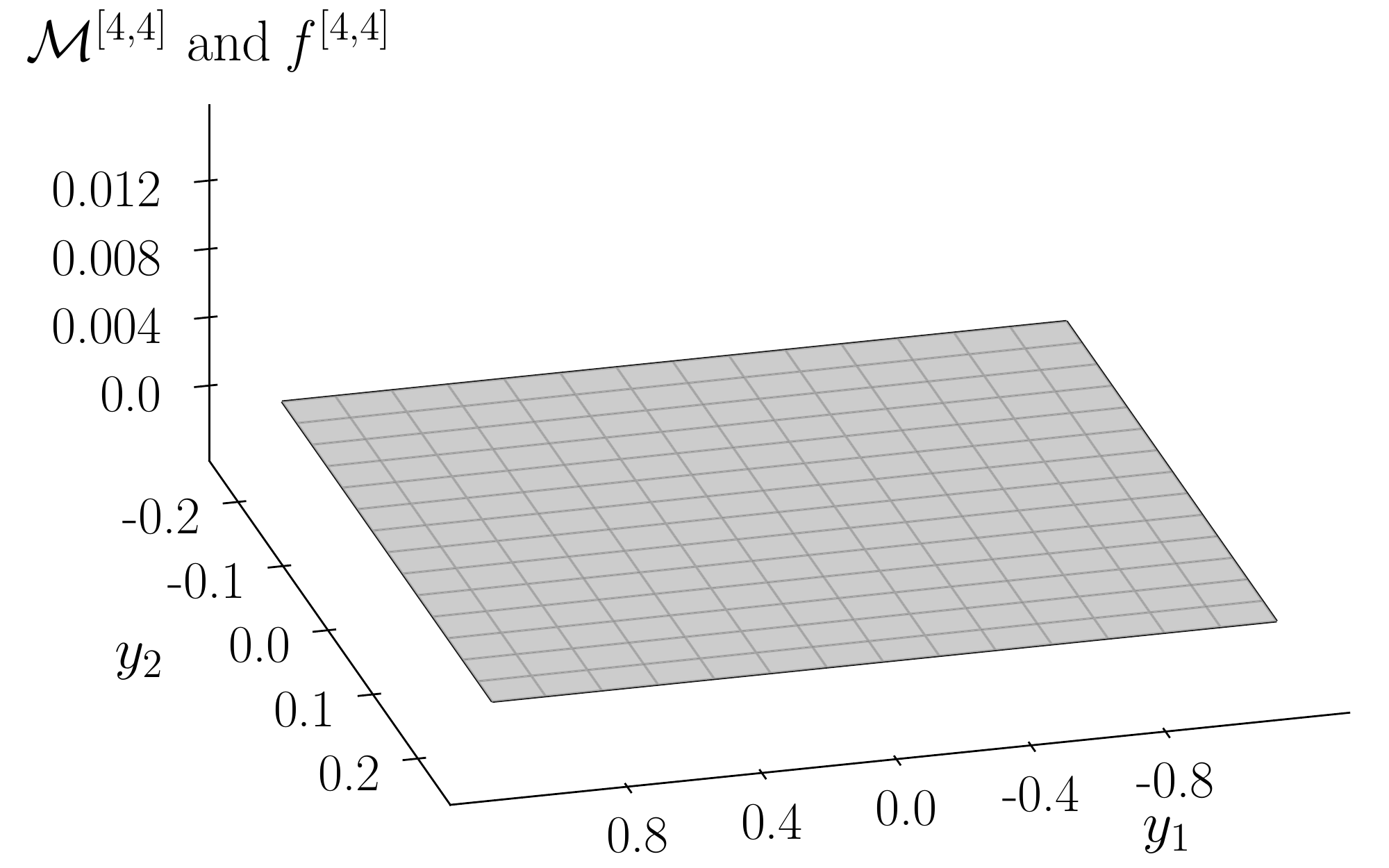}
\caption{Two-dimensional Kramers--Moyal (KM) coefficients $\Mtwo{\ell}{m}$ of bivariate diffusion processes given by Eq.~\eqref{eq:Cubib_JD_ms} (with $\phi=0.0$) together with the theoretically expected functions $\ftwo{\ell}{m}$ associated with each coefficient according to Eqs.~\eqref{eq:drift_km2}, \eqref{eq:diffusion_km2}, and \eqref{eq:jumps_km2}.
KM coefficients $\Mtwo{1}{0}$, $\Mtwo{0}{1}$, $\Mtwo{1}{1}$, $\Mtwo{2}{0}$,  $\Mtwo{0}{2}$,  $\Mtwo{2}{2}$,  $\Mtwo{4}{0}$,  $\Mtwo{0}{4}$, and $\Mtwo{4}{4}$ are shown.
Although seemingly small, the higher-order moments are all present and non-zero.
We find $\Mtwo{4}{0}=0.012$ and $\Mtwo{0}{4}=0.009$, as expected from Eq.~\eqref{eq:jumps_km2}.
All obtained KM surfaces fit considerably well their theoretically expected ones ($V_\text{err}^{[1,0]}= 0.68$, $V_\text{err}^{[0,1]}= 0.18$, $V_\text{err}^{[1,1]}< 0.01$, $V_\text{err}^{[2,0]}=0.01$, $V_\text{err}^{[0,2]}=0.01$, $V_\text{err}^{[2,2]}<0.01$, $V_\text{err}^{[4,0]}=0.01$, $V_\text{err}^{[0,4]}=0.01$,  $V_\text{err}^{[4,4]}<0.01$.; error volumes estimated over the displayed domain).}
\label{fig:Cubic_JD_simple}
\end{figure*}

Let $\ftwo{\ell}{m}(y_1,y_2)$ denote the theoretical value for orders $(\ell,m)$ introduced in the model, i.e., a non-linear combination of the various parameters of the system.
The distance between each surface can be defined as
\begin{equation}
\begin{aligned}\label{eq:least-square-volume}
\int\!\!\!\!\int_U\left(\Mtwo{l}{m}(y_1,y_2) - \ftwo{l}{m}(y_1,y_2)\right)^2 \d y_1 \d y_2 =: V^2,
\end{aligned}
\end{equation}
where $U$ denotes the domain of $\Mtwo{l}{m}(y_1,y_2)$.
The least-squared distance volume $V$ between the surfaces is zero if $\Mtwo{l}{m}(y_1,y_2)=\ftwo{l}{m}(y_1,y_2)$.
It is this volume that one aims to minimize such that the reconstructed KM coefficients match the underlying theoretical functions in the model.
Since $\Mtwo{l}{m}(y_1,y_2)$ is a real-valued function measured over a distribution space $U$, the density of data points is not uniform over $U$.
This implies that a comparative measure on distances between $\Mtwo{l}{m}(y_1,y_2)$ and $\ftwo{l}{m}(y_1,y_2)$ would be non-normalized to the density of points of the space.
We therefore introduce a normalization to  Eq.~\eqref{eq:least-square-volume} that ensures the less dense areas of $U$ are normalized accordingly, thus mitigating the effect of scarcity of points at the borders of $U$ and an overestimation of $V$ due to outliers in the distribution.
We derive such a normalization by considering the zeroth-order KM coefficient $\Mtwo{0}{0}(y_1,y_2)$ which captures exactly the density of points in $U$, although it is in itself not normalized as a distribution.
The resulting normalized volume error measure $V_\text{err}$ between surfaces takes the form (state dependencies not explicit)
\begin{equation}
\begin{aligned}\label{eq:least-square-error}
\int\!\!\!\!\int_U\left(\Mtwo{l}{m} - \ftwo{l}{m}\right)^2 p(y_1,y_2) \d y_1 \d y_2 = V_\text{err}^2,
\end{aligned}
\end{equation}
where $p(\cdot)$ denotes the probability density.
Coincidentally, the numerical evaluation implemented via both a histogram or a kernel-based estimators immediately yields this density. (i.e., the zeroth-power of the right-hand side of Eq.~\eqref{eq:2D_KMC}, before applying the estimation operator).
This makes it easy to retrieve $p(y_1,y_2)$ as one numerically evaluates data.

With this at hand, it is now possible to relate theoretical and numerical results and to quantify the deviation of the obtained KM coefficients from the functions employed.

To showcase what two-dimensional KM coefficients are as well as how to identify drift and diffusion terms of bivariate diffusion processes, we present in the following two exemplary processes with a priori known coefficients.
In this manner, by employing Eqs.~\eqref{eq:drift_km2_diff}~and~\eqref{eq:diffusion_km2_diff}, one can judge the outcome of the KM coefficient estimation procedure from discrete data in comparison with the expected theoretical functions.

We begin with two uncoupled processes, where one has constant diffusion and a quartic potential as the drift term:
\begin{equation}\label{eq:example_cubic_constant_noise}
\begin{aligned}
 &\vec{N}=\begin{pmatrix}
 N_1 \\ N_2
 \end{pmatrix} = \begin{pmatrix}
 -x_1^3 + x_1 \\ -x_2
 \end{pmatrix},\\
 &\vec{g} = \begin{pmatrix}
 g_{1,1} & g_{1,2} \\ g_{2,1} & g_{2,2}
 \end{pmatrix} = \begin{pmatrix}
 0.5 & 0.0 \\ 0.0  & 0.5
 \end{pmatrix},
 \end{aligned}
\end{equation}

In Fig.~\ref{fig:Cubic_constant_noise}, we show the corresponding KM coefficients $\Mtwo{1}{0}$, $\Mtwo{0}{1}$, $\Mtwo{1}{1}$, and $\Mtwo{2}{0}$ together with the theoretically expected functions.
The per-design cubic-linear function ($N_1=-x_1^3+x_1$) acting as drift coefficient along the first dimension as well as the negatively-sloped surface of $N_2 = -x_2$ are evident.
Likewise, the constant diffusion term leads to a flat constant-valued $\Mtwo{2}{0}$, and the absence of any non-diagonal elements ($g_{1,2}=g_{2,1}=0$) agrees with the zero-valued $\Mtwo{1}{1}$.
Alongside the surfaces are plotted the theoretically expected values, which agree well with the data recovery.

We next extent Eq.~\eqref{eq:example_cubic_constant_noise} by adding multiplicative noise to the diffusion term and by including a noise coupling term $g_{1,2}\neq 0$
\begin{equation}\label{eq:example_cubic_weird_noise}
\begin{aligned}
 &\vec{N}=\begin{pmatrix}
 N_1 \\ N_2
 \end{pmatrix} = \begin{pmatrix}
 -x_1^3 + x_1 \\ -x_2
 \end{pmatrix},\\
 &\vec{g} = \begin{pmatrix}
 g_{1,1} & g_{1,2} \\ g_{2,1} & g_{2,2}
 \end{pmatrix} = \begin{pmatrix}
 0.1 + x_1^2 & 0.5 \\ 0.0  & 0.2 + 2x_2^2
 \end{pmatrix}.
 \end{aligned}
\end{equation}
The recovered KM coefficients (see Fig.~\ref{fig:Cubic_weird_noise}) of the drift terms remain unaltered, but as posited, the second-order KM coefficients, i.e., $\Mtwo{1}{1}$, $\Mtwo{2}{0}$, $\Mtwo{0}{2}$, and $\Mtwo{2}{2}$, clearly exhibit the influence of the multiplicative noise.
The quadratic multiplicative dependencies of $\Mtwo{2}{0}$ and $\Mtwo{0}{2}$ and their offsets from zero are evident.
More pertinently, one can notice $\Mtwo{1}{1}$ to display the expected shape arising from Eq.~\eqref{eq:diffusion_km2_diff}, i.e., this value is non-zero and exhibits the parabolic shape of $g_{1,2}\cdot g_{2,2} = 0.5 (0.2 + 2x_2^2)$.
For $x_2=0$, the minimum of $\Mtwo{1}{1}$ coincides with $0.1$, as expected.
This indicates that the presence of the multiplicative noise does not hinder the assertion of the KM coefficients.
Again, the recovered KM coefficients match the theoretically ones.

\section{Bivariate jump-diffusion processes}
The KM coefficients of bivariate jump-diffusion processes take the following form (under the parameter prescription used in the jump-diffusion model in Eq.~\eqref{eq:model})~\cite{anvari2016,lehnertz2018,Tabar-book2019}
\begin{align}
&\begin{aligned}\label{eq:drift_km2}
\Mtwo{1}{0} &=\, N_1\\
\Mtwo{0}{1} &=\, N_2
\end{aligned}\\[1em]
&\begin{aligned}\label{eq:diffusion_km2}
\Mtwo{1}{1} &= \,g_{1,1} g_{2,1} ~+~ g_{1,2} g_{2,2} \\
\Mtwo{2}{0} &= \left[\,g_{1,1}^2 + s_{1,1} \lambda_1 ~+~ g_{1,2}^2 + s_{1,2} \lambda_2\right] \\
\Mtwo{0}{2} &=\, \left[g_{2,1}^2 + s_{2,1} \lambda_1 ~+~ g_{2,2}^2 + s_{2,2} \lambda_2\right]
\end{aligned}\\[1em]
&\begin{aligned}\label{eq:jumps_km2}
\Mtwo{2}{2} &= \,~\, \left[s_{1,1} s_{2,1} \lambda_1 ~+~ s_{1,2} s_{2,2} \lambda_2 \right]\\
\Mtwo{4}{0} &= \,~3~\, \left[s_{1,1}^2 \lambda_1 ~+~ s_{1,2}^2 \lambda_2\right] \\
\Mtwo{0}{4} &= \,~3~\, \left[s_{2,1}^2 \lambda_1 ~+~ s_{2,2}^2 \lambda_2\right] \\
\Mtwo{4}{4} &= \, ~9~ \left[s_{1,1}^2 s_{2,1}^2 \lambda_1 ~+~ s_{1,2}^2 s_{2,2}^2 \lambda_2\right]
\end{aligned}\\[1em]
&\begin{aligned}
\Mtwo{6}{0} &= \,~15~  \left[s_{1,1}^3 \lambda_1 ~+~ s_{1,2}^3 \lambda_2\right] \\
\Mtwo{0}{6} &= \,~{15}~ \left[s_{2,1}^3 \lambda_1 ~+~ s_{2,2}^3 \lambda_2\right] \\
\Mtwo{6}{6} &= \, ~{225}~ \left[s_{1,1}^3 s_{2,1}^3 \lambda_1 ~+~ s_{1,2}^3 s_{2,2}^3 \lambda_2\right]
\end{aligned}
\end{align}
\begin{align}
&\begin{aligned}
\Mtwo{8}{0} &= \,~{105}~  \left[s_{1,1}^4 \lambda_1 ~+~ s_{1,2}^4 \lambda_2\right] \\
\Mtwo{0}{8} &= \,~{105}~  \left[s_{2,1}^4 \lambda_1 ~+~ s_{2,2}^4 \lambda_2\right]\\
\Mtwo{8}{8} &= \, ~{11025}~  \left[s_{2,1}^4 \lambda_1 ~+~ s_{2,2}^4 \lambda_2\right]
\end{aligned}
\end{align}
\begin{figure*}[t]
\includegraphics[width=.32\linewidth]{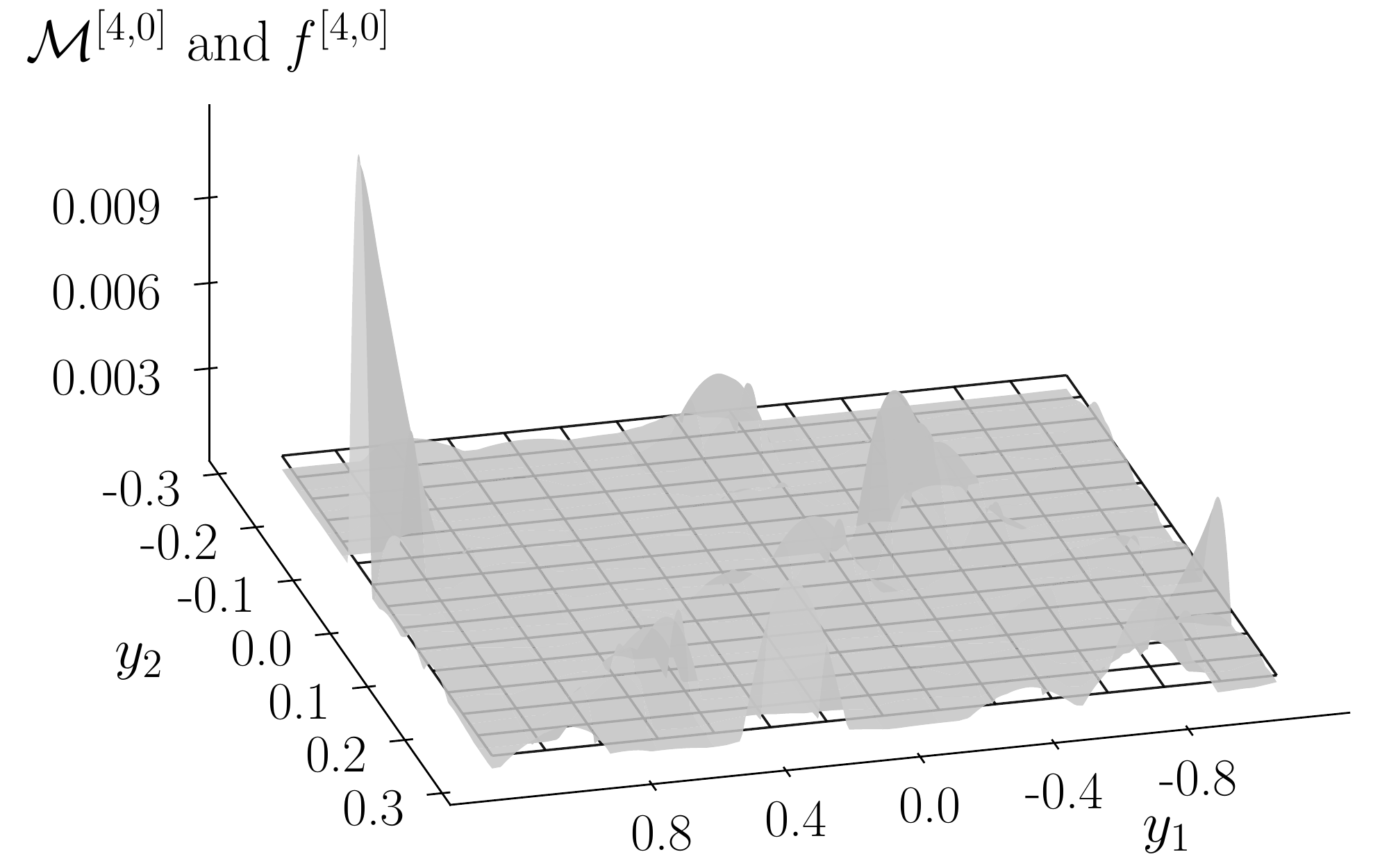}
\includegraphics[width=.32\linewidth]{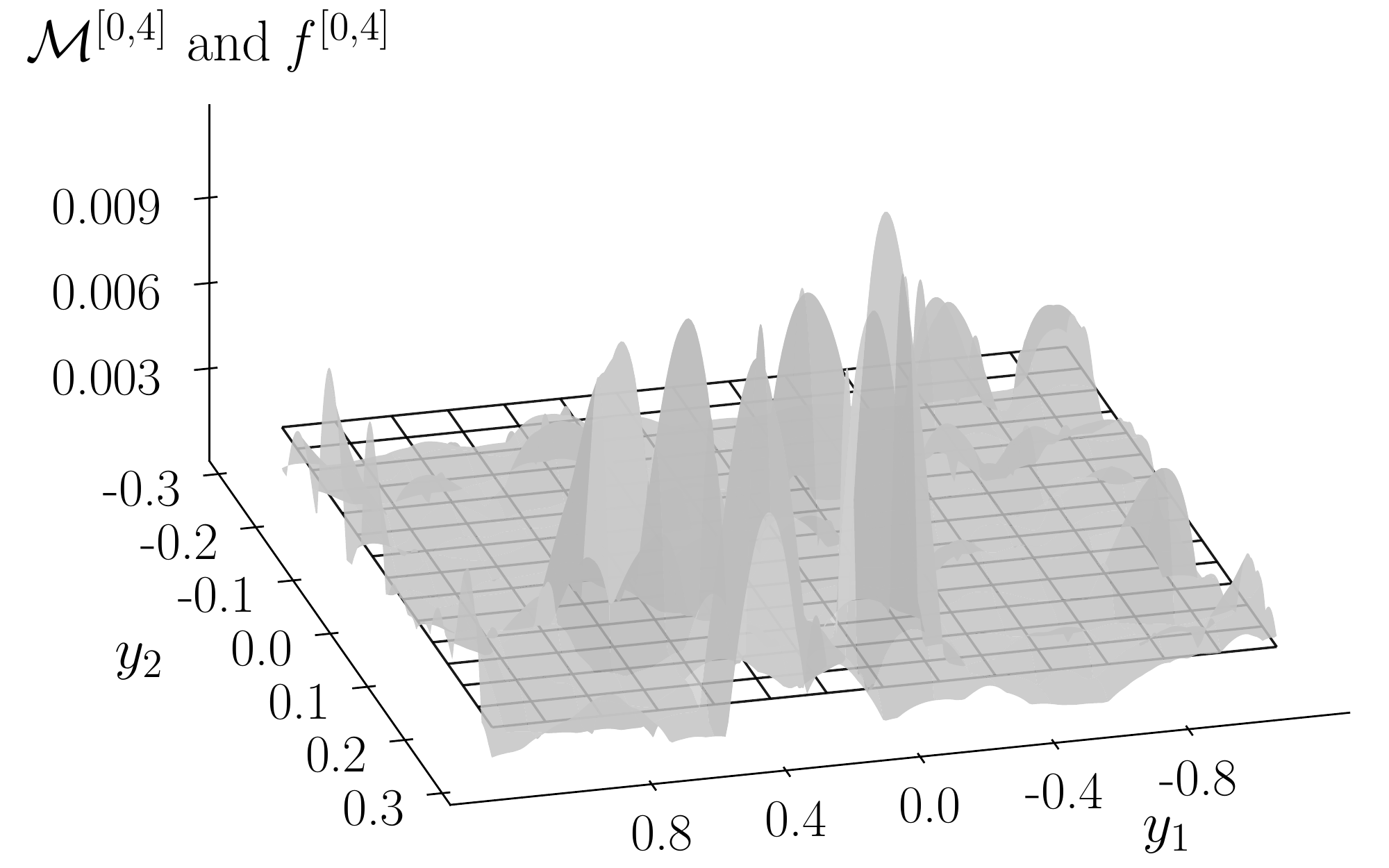}
\includegraphics[width=.32\linewidth]{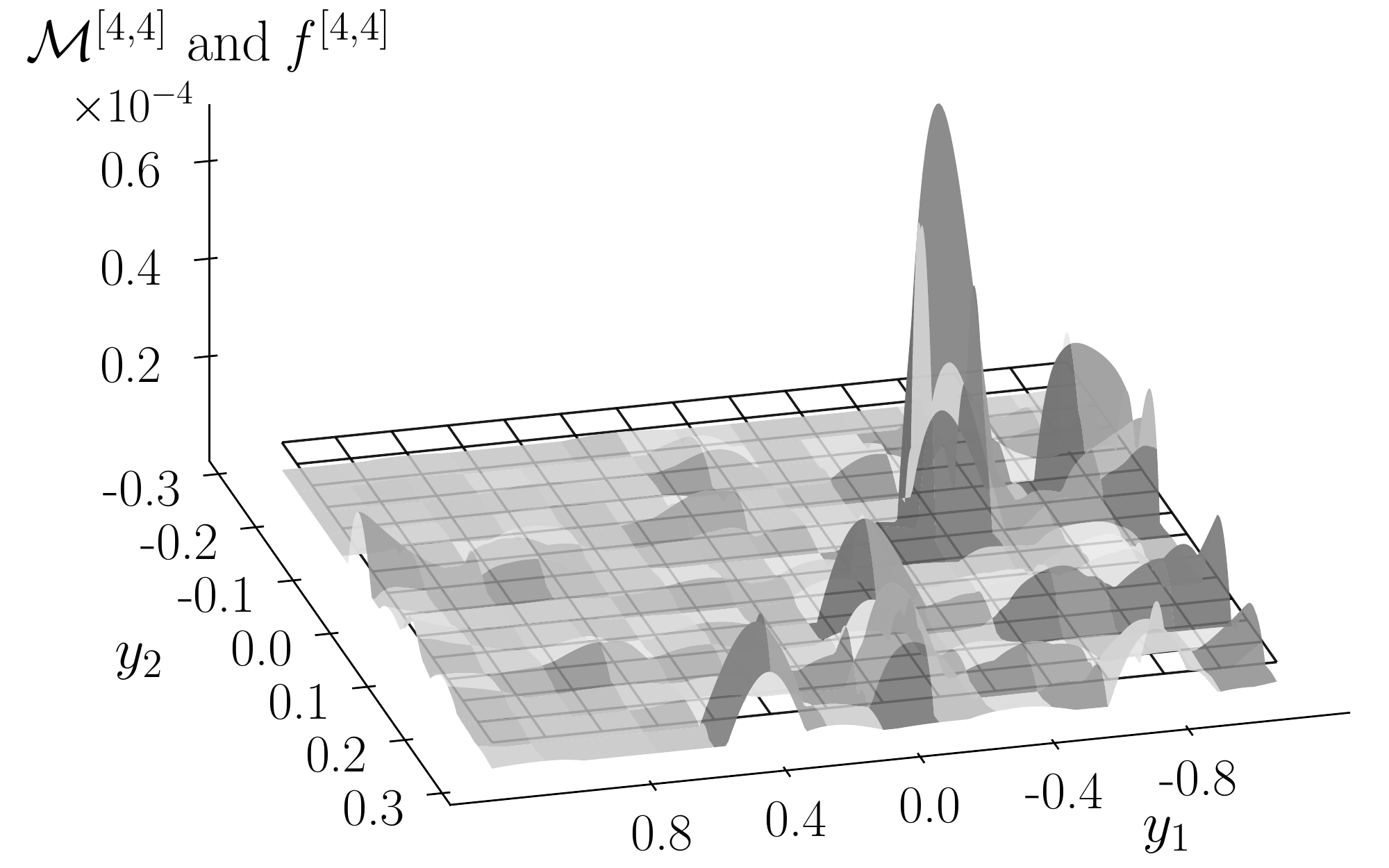}
\caption{Two-dimensional Kramers--Moyal (KM) coefficients $\Mtwo{4}{0}$, $\Mtwo{0}{4}$, and $\Mtwo{4}{4}$ of bivariate jump-diffusion processes given by Eq.~\eqref{eq:Cubib_JD_ms} (with $\phi=0.3$)
together with the respective theoretically expected functions $f$, associated with each coefficient according to Eqs.~\eqref{eq:drift_km2}, \eqref{eq:diffusion_km2}, and \eqref{eq:jumps_km2}.
Notice that the estimated KM coefficients agree well with the theoretical expected functions in all orders (error volumes estimated over the displayed domain).
For further details, see Appendix~\ref{App:Cubib_JD_ms}.}
\label{fig:Cubic_JD_ms_inline}
\end{figure*}
where $\langle \xi_{ij}^{2\ell} \rangle = s_{ij}^\ell$ are the variances of the Gaussian-distributed jump amplitudes.
An extended derivation can be found in Appendix~\ref{App:1}.
The last equations here are taken from the general form
\begin{equation}\label{eq:higher_2D_KMC}
\begin{aligned}
M^{[2\ell,2m]} = \frac{(2\ell)!}{2^\ell \ell!} \frac{(2m)!}{2^m m!}  \left[  s_{1,1}^{\ell} s_{2,1}^{m}  \lambda_1 +
 s_{1,2}^{\ell} s_{2,2}^{m}  \lambda_2 \right]. \nn
\end{aligned}
\end{equation}

\subsection{Understanding the impact of jumps}
As an illustrative case study, we investigate a general jump-diffusion process that is based on Eq.~\eqref{eq:example_cubic_weird_noise} but excludes the multiplicative diffusion terms.
Taking into account the effect of the jump terms, but maintaining the system independent in at least one of the dimensions, we extend Eq.~\eqref{eq:example_cubic_weird_noise} to include jumps only in the diagonal terms of $\vec{\xi}$
\begin{equation}\label{eq:Cubib_JD_ms}
\begin{aligned}
 &\vec{N}=\begin{pmatrix}
 N_1 \\ N_2
 \end{pmatrix} = \begin{pmatrix}
 -x_1^3 + x_1 \\ -x_2
 \end{pmatrix}\\
 &\vec{g} = \begin{pmatrix}
 g_{1,1} & g_{1,2} \\ g_{2,1} & g_{2,2}
 \end{pmatrix} = \begin{pmatrix}
 0.1 & 0.5 \\ 0.0 & 0.2
 \end{pmatrix}\\
 &\vec{\xi} = \begin{pmatrix}
 \xi_{1,1} & \xi_{1,2} \\ \xi_{2,1} & \xi_{2,2}
 \end{pmatrix} = \begin{pmatrix}
 0.2 & 0.0 \\ \phi & 0.1
 \end{pmatrix}\\
 &\vec{\lambda} =\begin{pmatrix}
 \lambda_1 \\ \lambda_2
 \end{pmatrix} = \begin{pmatrix}
 0.1 \\ 0.3
 \end{pmatrix},
 \end{aligned}
\end{equation}
where for the present case, $\phi=0.0$.
In this manner, jumps are added to the first dimension of the process, having an amplitude of $ \xi_{1,1} = 0.2$ and occurring every $0.1 t$, given $\lambda_1 = 0.1$.
Similarly, jumps are added to the second dimension, $\xi_{2,2} = 0.1$, but the jumps occur three times more often than the aforementioned, given $\lambda_2=0.3$.
%
The influence of jumps can be observed across all KM coefficients (see Fig.~\ref{fig:Cubic_JD_simple}).
The previously smooth KM surfaces become rugged from the fast variations emerging due to the jumps, and the higher-order KM coefficients---although small compared to the lower-order ones---clearly do not vanish.
This indicates that the continuous stochastic modeling of time series of complex systems (white-noise-driven Langevin equation) together with the commonly used assumptions stemming from the Pawula theorem are invalid for jump-diffusion processes.
Modeling these processes with only the first two orders of the KM expansion of the master equation is therefore insufficient.

In order to understand further if it is possible to uncover the jump amplitude terms of coupled processes, we use the previous model Eq.~\eqref{eq:Cubib_JD_ms} with $\phi=0.3$, thereby effectively introducing a coupling via the off-diagonal elements of the jump matrix $\vec{\xi}$.
We show, in Fig~\ref{fig:Cubic_JD_ms_inline}, the corresponding fourth-order KM coefficients.
The impact of the coupling is visible, although small, in  $\Mtwo{4}{4}$, which is no longer zero.
Likewise, $\Mtwo{4}{0}$ and $\Mtwo{0}{4}$ also do not vanish.
In Appendix~\ref{App:Cubib_JD_ms}, we present the corresponding KM coefficients up to order eight.

\subsection{Criteria for recovering coefficients in diffusion and jump-diffusion models}
For the case of vanishing off-diagonal elements $g_{2,1}$ and $\xi_{1,2}$, we can identify ways to recover the remaining coefficients of these processes.

First, given that the noise $\vec{\d} \vec{\omega}$ is Gaussian distributed, $\vec{g}$ is sign-reversal symmetric and one can thus assume that it takes only positive values.
One obtains that, if $\Mtwo{1}{1}=0$ then at least two elements of $\vec{g}$ must be zero, and if $\Mtwo{2}{2}=0$ then at least two elements of $\vec{\xi}$ must be zero (by assuming that $\lambda_1$ and $\lambda_2$ are non-vanishing rates).
These findings reduce the dimensionality of the estimation procedure and ensure that the underlying processes are less complex that the full-fledged description of Eq.~\eqref{eq:model}, although they do not grant which coefficients are zero-valued.

Second, if one either employs a heuristic argument of independence of the jump processes or neglects the off-diagonal jump amplitudes $\xi_{1,2}$ and $\xi_{2,1}$ (e.g., by assuming they are small compared to the diagonal terms of $\vec{\xi}$), one finds the following approximations:
\begin{equation}\label{eq:approx_jumps}
\begin{aligned}
\frac{1}{5}\frac{\Mtwo{6}{0}}{\Mtwo{4}{0}} = \frac{1}{5}\frac{15}{3}\frac{s_{1,1}^3\lambda_1}{s_{1,1}^2\lambda_1} &= s_{1,1}, \\
\frac{1}{5}\frac{\Mtwo{0}{6}}{\Mtwo{0}{4}} = \frac{1}{5}\frac{15}{3}\frac{s_{2,2}^3\lambda_1}{s_{2,2}^2\lambda_1}&=s_{2,2}.
\end{aligned}
\end{equation}
Likewise, the jump rates $\lambda_1$ and $\lambda_2$ can be obtained equivalently as
\begin{equation}\label{eq:approx_jumps_2}
\begin{aligned}
\frac{105}{9}\frac{{\Mtwo{4}{0}}^2}{\Mtwo{8}{0}} = \frac{105}{9}\frac{3^2}{105}\frac{(s_{1,1}^2\lambda_1)^2}{s_{1,1}^4\lambda_1} &= \lambda_1, \\
\frac{105}{9}\frac{{\Mtwo{0}{4}}^2}{\Mtwo{0}{8}} = \frac{105}{9}\frac{3^2}{105}\frac{(s_{2,2}^2\lambda_2)^2}{s_{2,2}^4\lambda_2} &= \lambda_2.
\end{aligned}
\end{equation}
Taking again model Eq.~\eqref{eq:Cubib_JD_ms} with $\phi=0.0$ and following Eq.~\eqref{eq:approx_jumps}, we obtain
\begin{equation}
\begin{aligned}
	& s^{\mathrm{est}}_{1,1} = 0.16 \approx 0.2 = s_{1,1} , \nn \\
	& s^{\mathrm{est}}_{2,2} = 0.09 \approx 0.1 = s_{2,2}. \nn
\end{aligned}
\end{equation}
These estimated values (indicated by the superscript $\cdot^\mathrm{est}$) are very close to the actual ones.
The criteria and approximations are especially relevant when constructing or analyzing systems which are known to have a specific unidirectional coupling form, e.g. a master-slave system, where for example the noise or the slave system are dictated by the driving master system.

\begin{figure}[t]
\includegraphics[width=1\linewidth]{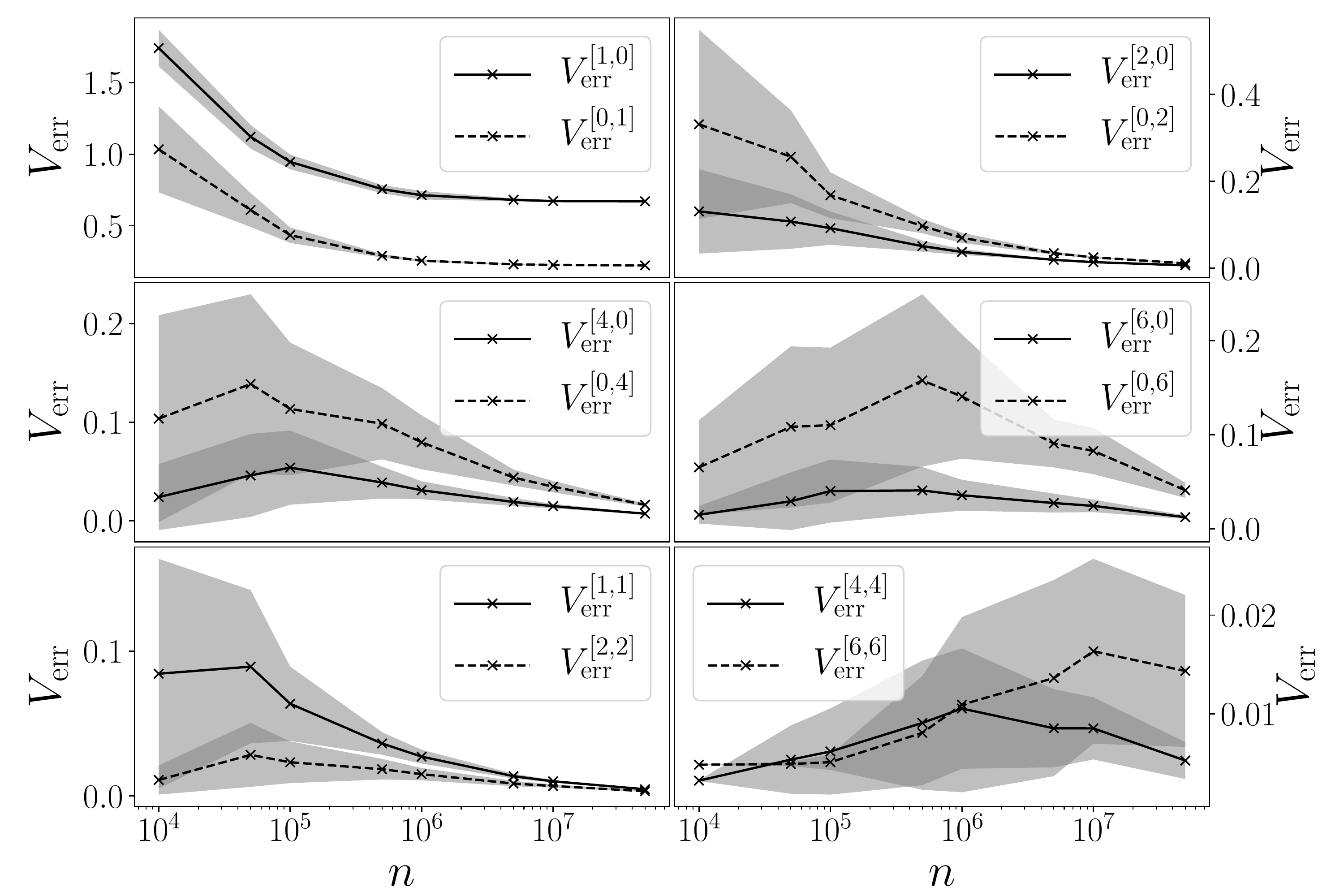}
\caption{[Abscissa in logarithmic scale] Error volume $V_\text{err}$ for a bivariate jump-diffusion process (Eq.~\eqref{eq:Cubib_JD_ms}) depending on the number of data points $n$ in the time series.
Each process is numerically integrated with random initial conditions, for varying number of data points $n\in[10^3, 10^8]$ and over $50$ times.
The time sampling used was of $10^{-3}$.
The average value of $V_\text{err}$ and one standard deviation (shaded area) are displayed.
Notice the clear decrease on all KM coefficients with either $\ell=0$ or $m=0$, e.g. $\Mtwo{2}{0}$ or $\Mtwo{0}{2}$, as the number of data points $n$ increases.
This can be seen since the volume between the theoretically expected values and the KM coefficients decreases consistently, i.e., $V_{\text{err}}^{[\ell,m]}$ decreases for an increasing number of data points.
The KM coefficients with $\ell\neq 0$ and $m\neq 0$, such as $\Mtwo{4}{4}$ or $\Mtwo{6}{6}$ present themselves as non-decreasing, but the error volume is overall considerable small in value (cf. Fig.~\ref{fig:No_points_complex}).
It is important to notice that $V_{\text{err}}^{[1,1]}$ does not converge to zero since the KM coefficient is associated with the quartic potential (i.e., the term $N_1=-x^3+x$).
Due to its shape, the process has two preferred states, either at $x=-1$ or $x=1$, thus spends little time at any intermediary point, like $x=0$, damaging the statistics of the recovery.}
\label{fig:No_points_ms}
\end{figure}

\subsection{Factors influencing quality of recovery of coefficients}
In order to validate the quality of the non-parametric recovery of the KM coefficients, we now turn to two critical aspects:
firstly, bivariate processes may require a high number of data points in a time series for the estimation to be reliable;
secondly, the interplay between the drift, diffusion, and jump parts of a stochastic processes may render the estimation incorrect.

Addressing these aspects, we include a more contrived model involving couplings and interactions in both the diffusion and jump terms, thus theoretically resulting in having all higher-order KM coefficients non-zero, and especially the KM coefficients with $\ell\neq 0$ and $m\neq 0$.
The parameters for the model read:
\begin{equation}\label{eq:Cubib_JD_complex}
\begin{aligned}
 &\vec{N}=\begin{pmatrix}
 N_1 \\ N_2
 \end{pmatrix} = \begin{pmatrix}
 -x_1^3 + x_1 \\ -x_2
 \end{pmatrix},\\
 &\vec{g} = \begin{pmatrix}
 g_{1,1} & g_{1,2} \\ g_{2,1} & g_{2,2}
 \end{pmatrix} = \begin{pmatrix}
 0.1 & 0.5 \\ \alpha  & 0.2
 \end{pmatrix},\\
 &\vec{\xi} = \begin{pmatrix}
 \xi_{1,1} & \xi_{1,2} \\ \xi_{2,1} & \xi_{2,2}
 \end{pmatrix} = \begin{pmatrix}
 0.2 & 0.5 \\ \beta & 0.1
 \end{pmatrix},\\
 &\vec{\lambda} =\begin{pmatrix}
 \lambda_1 \\ \lambda_2
 \end{pmatrix} = \begin{pmatrix}
 0.1 \\ 0.3
 \end{pmatrix}.\\
 \end{aligned}
\end{equation}
The parameters $\alpha$ (diffusion term) and $\beta$ (jump term) can be freely varied.

\begin{figure}[t]
\includegraphics[width=1\linewidth]{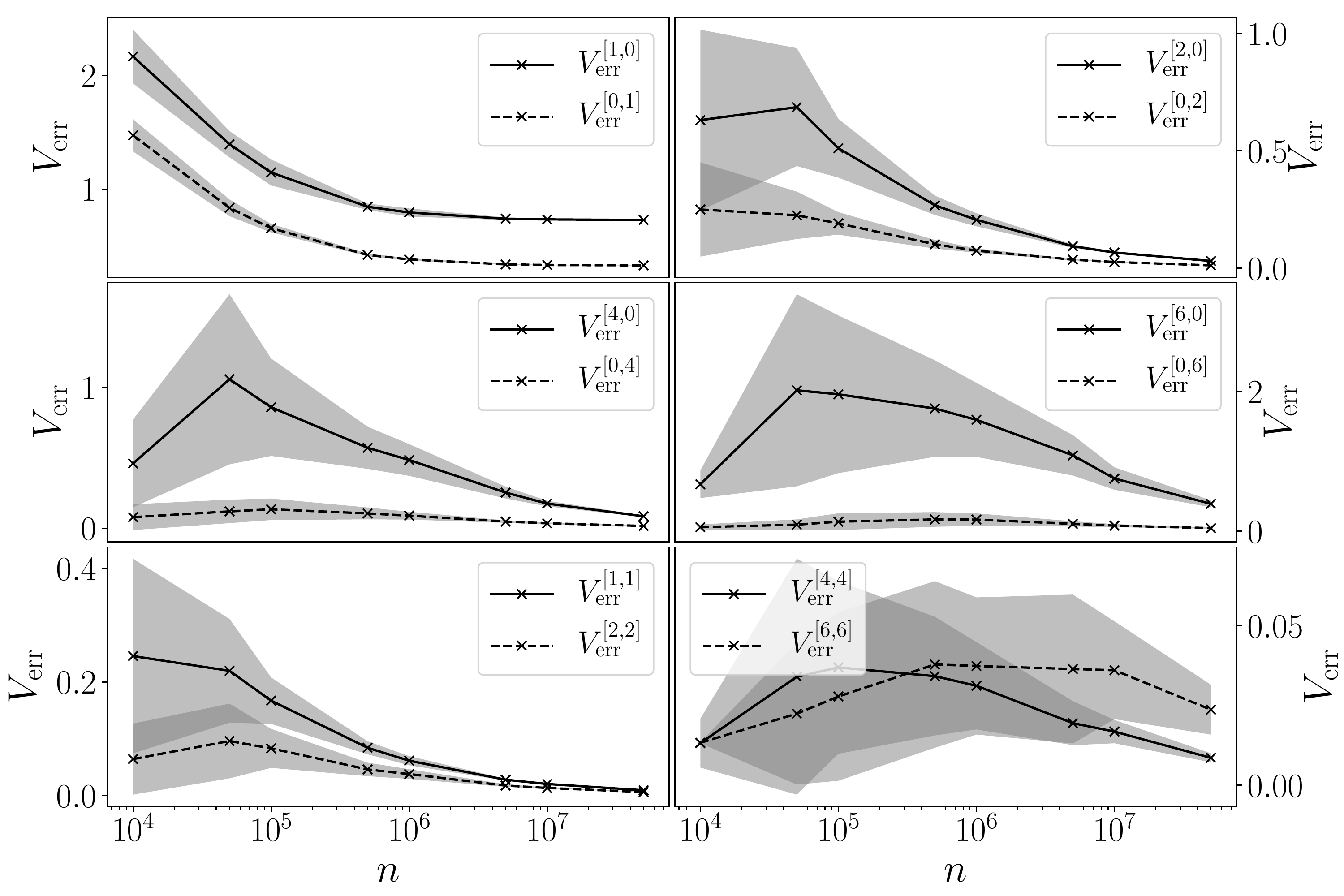}
\caption{[Abscissa in logarithmic scale] Same as Fig.~\ref{fig:No_points_ms} but for the bivariate jump-diffusion process (Eq.~\eqref{eq:Cubib_JD_complex}) with $\alpha=\beta=0.3$.
Integration parameters as in Fig.~\ref{fig:No_points_ms}.}\label{fig:No_points_complex}
\end{figure}

Let us focus firstly on the number of data points in a time series.
We utilize the models Eq.~\eqref{eq:Cubib_JD_ms} with $\phi=0.3$ and Eq.~\eqref{eq:Cubib_JD_complex} with $\alpha=\beta=0.3$, and show in Figs.~\ref{fig:No_points_ms} and~\ref{fig:No_points_complex},
respectively, the error volumes $V_\text{err}$ for the KM coefficients for an increasing number of data points.
The reliability of the recovery of KM coefficient is valid for higher amount of data ($n \geq 10^5$), as expected, although the scarcity of data posits no extensive problem for the calculation.
It is especially important to notice that a time series with a lower amount of data entails naturally fewer jumps in the process, hindering the possibility of accurately recovering the jump terms from such short time series.
For $n\geq 10^6$, 
the estimation seems reliable, the standard deviations become minute, and most error values approach zero, i.e., the theoretical and estimated KM surfaces are close.

One remark is necessary on the recovery of the drift terms.
The presence of noise and jumps in the process takes its toll on the recovery of the exact form of the KM coefficients as well as the explicit dependence of the state variables, i.e., the quartic potential in both Eqs.~\eqref{eq:Cubib_JD_ms} and~\eqref{eq:Cubib_JD_complex}.
A finer time sampling can help to improve the results.

\begin{figure}[t]
\includegraphics[width=1\linewidth]{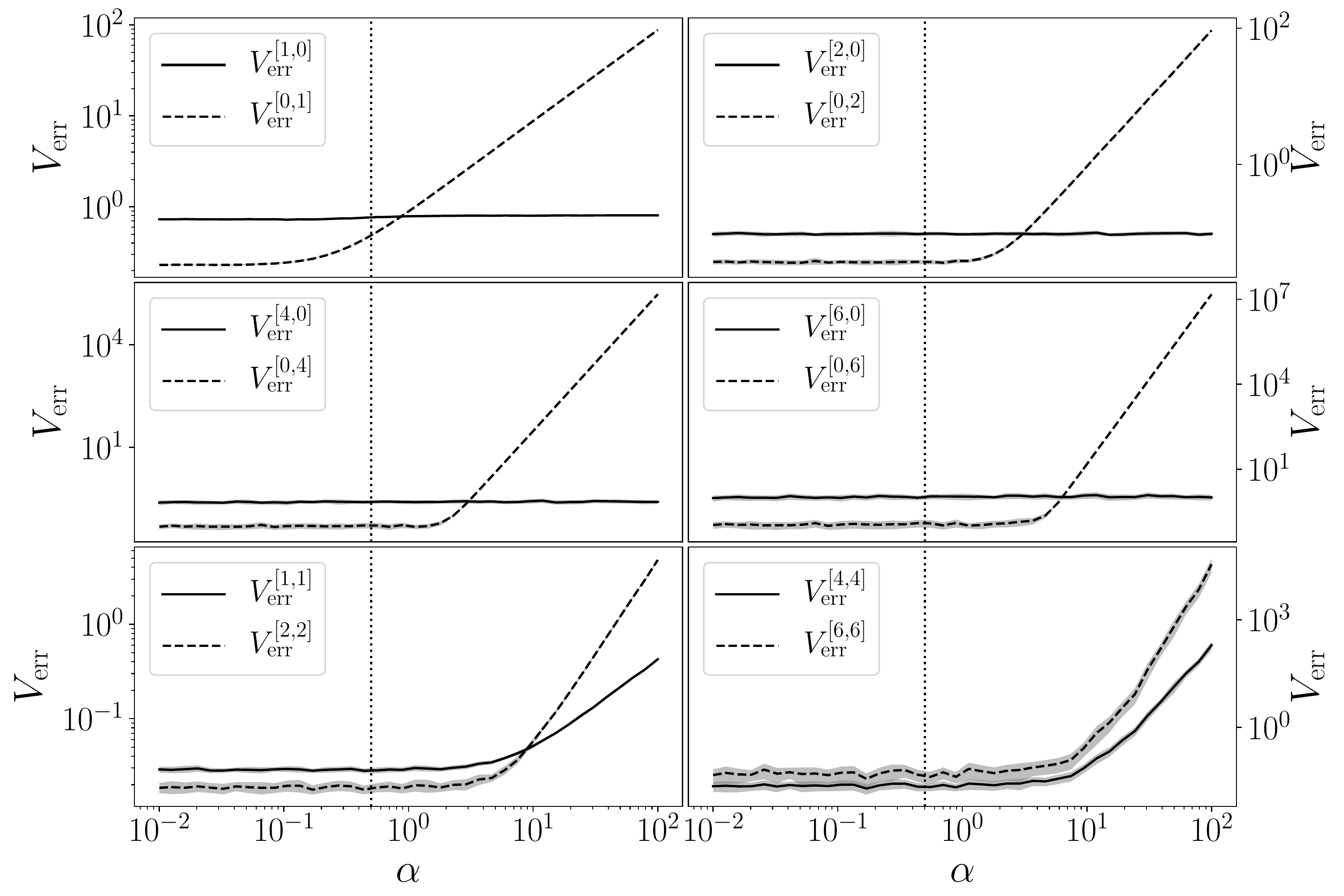}
\caption{[Double logarithmic scale] Error volume $V_\text{err}$ for the bivariate jump-diffusion process Eq.~\eqref{eq:Cubib_JD_complex} for a varying diffusion term $\alpha \in [10^{-2},10^{2}]$.
The vertical dotted line at $\alpha=0.5$ indicates the point where the diffusion term $\alpha = g_{2,1} = 0.5$ is equal to $g_{1,2} = 0.5$.
A small value of the diffusion term $\alpha$, in comparison to the  diffusion parameters $g_{2,1}$ and $g_{1,2}$, ensures a good reconstruction, i.e., a small error volume $V_\text{err}$.
The average and one-standard deviations (shaded area) are displayed.
For each point $50$ iterations are taken, each with a total number of data points of $5\times10^6$ and a time sampling of $10^{-3}$.}\label{fig:alpha}
\end{figure}

\begin{figure}[t]
\includegraphics[width=1\linewidth]{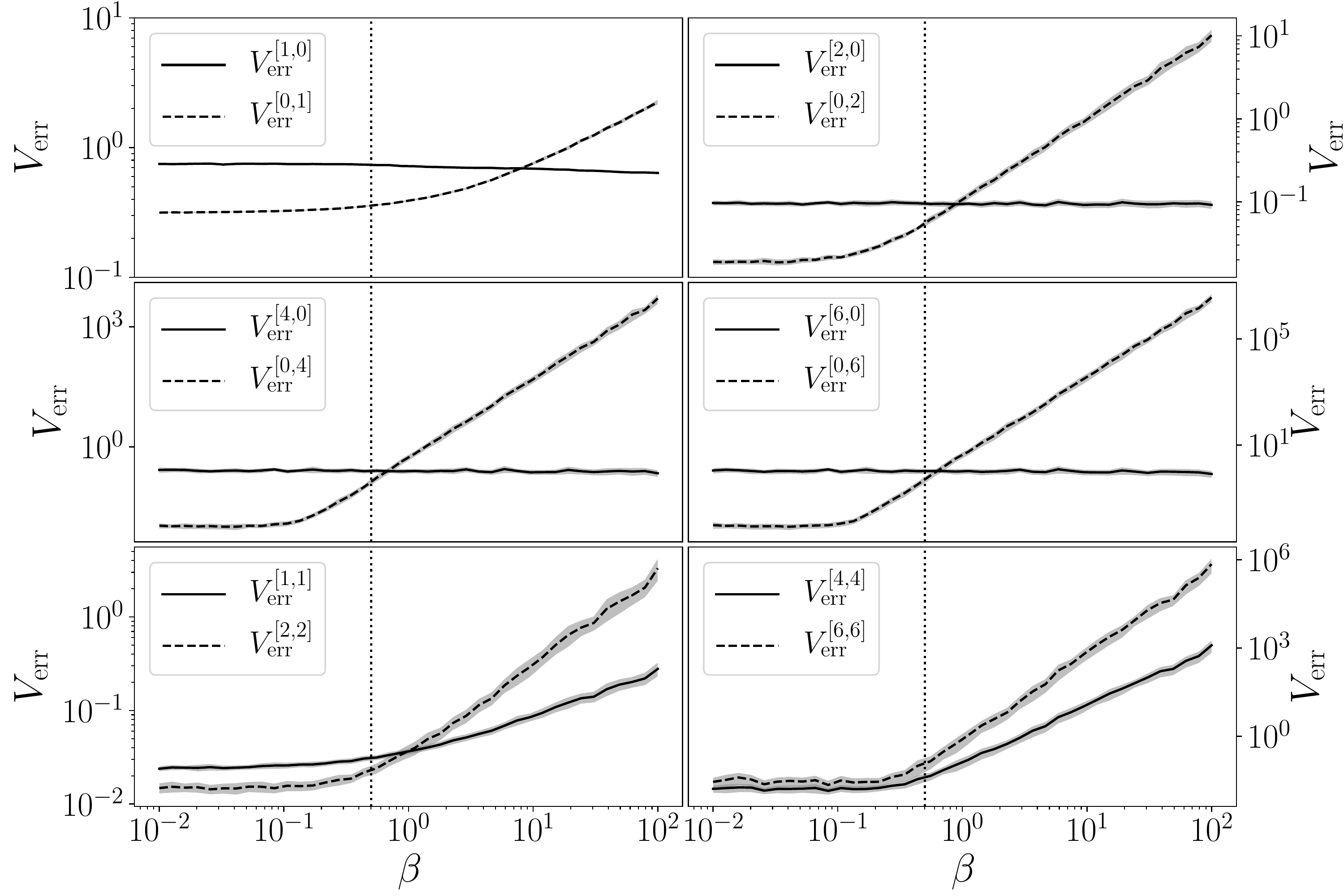}
\caption{[Double logarithmic scale] Error volume $V_\text{err}$ for the bivariate jump-diffusion process Eq.~\eqref{eq:Cubib_JD_complex} for a varying jump term $\beta \in [10^{-2},10^{2}]$.
The vertical dotted line indicates the biggest jump term value $\xi_{1,2}=0.5$ to compare with $\beta$.
As in direct analogy to Fig.~\ref{fig:alpha}, a small jump term $\beta$ ensures a good reconstruction, i.e., a small error volume $V_\text{err}$.
Increasing values of the jump term $\beta$ in comparison to the other parameters in the system make the reconstruction unreliable.
Iteration scheme identical to the one in Fig.~\ref{fig:alpha}.}\label{fig:beta}
\end{figure}

To further test the limitation of retrieving the KM coefficients from data, we utilize model Eq.~\eqref{eq:Cubib_JD_complex} once more and investigate the influence of parameters $\alpha$ (diffusion term) and $\beta$ (jump term).
For increasing $\alpha$ ($\alpha \in [10^{-2},10^{2}]$) and $\beta=0.3$,
we observe a considerable impact on the error volume $V_\text{err}$
after the order of magnitude on the diffusion parameter $\alpha$ is ten-fold bigger in comparison to the diffusion parameter $g_{1,2}$ (Fig.~\ref{fig:alpha}).
Similarly, for increasing $\beta$ ($\beta \in [10^{-2},10^{2}]$) and $\alpha=0.3$, the error volume $V_\text{err}$ is considerably impacted already when $\beta$ is of similar size as the other parameters, namely $\xi_{1,2}=0.5$ (Fig.~\ref{fig:beta}).

These findings point to the difficulty of recovering the KM coefficients in the presence of jumps.
Nonetheless, our findings indicate that the current understanding, modeling, and numerical recovery of KM surfaces, for the case of jumps of comparable size to the diffusion terms, is possible and reliable.

Before closing this section, we note that the calculations of the KM coefficients are computationally inexpensive~\cite{RydinGorjao2019}.
E.g., estimating all the 14 KM coefficients, $\Mtwo{1}{0}, \Mtwo{0}{1}, \dots, \Mtwo{0}{8}$ as presented in Fig.~\ref{fig:Cubic_JD_ms} takes about \unit[5.5]{s} on a desktop computer (quad-core with clock speed \unit[2.20]{GHz}) for a bi-dimensional time series of size $2\times10^6$ data points (and \unit[22]{s} for $2\times10^7$ data points).
Our approach might thus be advantageous for field applications that aim at an investigation of interactions between complex systems with poorly understood dynamics.

\section{Conclusion}

We introduced the bivariate jump-diffusion process which comprises two-dimensional diffusion and two-dimensional jumps that can be coupled to one another.
For such a process we presented a data-driven, non-parametric estimation procedure of higher-order Kramers--Moyal coefficients and investigated its pros and cons using synthetic bivariate time series from continuous and discontinuous processes.
The procedure allows one to reconstruct relevant aspects of the underlying jump-diffusion processes and to recover the underlying parameters.

Having now a traceable mathematical framework, the model can be extended to embody other noise and jump properties.
An extension from the underlying Wiener process to include e.g. fractional Brownian motion is straightforward but nevertheless requires further investigations to derive an explicit forward Kolmogorov equation \cite{Tabar-book2019}.
Also, a generalization to continuous jump processes---originating from alpha-stable or other heavy-tailed distributions (L\'evy noise-driven Langevin equation)---is possible, however, with the drawback that calculating the conditional moments may not always be mathematically possible \cite{Tabar-book2019}.
On the other hand, a numerical estimation of generalised moments should be possible but these still require a physical interpretation.

We are confident that our novel approach provides a general avenue to further understanding of interacting complex systems (e.g. brain or power grids~\cite{lehnertz2018,schafer2018,Gorjao2019,Anvari2019}) whose dynamics exhibit nontrivial noise contributions.

\section*{Acknowledgements}
The authors would like to thank Thorsten Rings and Mehrnaz Anvari for interesting discussions, and Francisco Meirinhos for the tremendous help in devising the methodology behind the results.
L.R.G. thanks Dirk Witthaut and Giulia di Nunno for everlasting support, and gratefully acknowledges support by the grant No. VH-NG-1025 and a scholarship from the E.ON Stipendienfonds.

~

\bibstyle{unsrt}
\bibliography{bib}

\pagebreak
\newpage~
\newpage
\pagebreak
\onecolumngrid

\begin{appendices}

\section{Extended derivation of the two-dimensional Kramers--Moyal coefficients for a jump-diffusion process}\label{App:1}

The following derivations stem from Eq.~\eqref{eq:2D_KMC} and apply to the two-dimensional jump-diffusion process $(y_1,y_2)$, as in Eq.~\eqref{eq:model}.
All orders of the Kramers--Moyal coefficients  $\Mtwo{\ell}{m}$ are $(\ell, m)\in\mathbb{N}_+$.

\subsubsection{Kramers--Moyal coefficients $\Mtwo{1}{0}$ and $\Mtwo{0}{1}$}
\begin{align*}
\Mtwo{1}{0}(x_1,x_2) &=  \lim_{\d t \to 0} \frac{1}{\d t} \langle (\d y_1)^{1} (\d x_2)^{0} \rangle|_{y_1(t)=x_1, y_2(t)=x_2} \\
            &= \lim_{\d t \to 0} \frac{1}{\d t} \langle \d y_1 \rangle|_{y_1(t)=x_1, y_2(t)=x_2} \\
            &=  \lim_{\d t \to 0} \frac{1}{\d t} \langle N_1 \d t + g_{1,1} \d w_1 + g_{1,2} \d w_2 + \xi_{1,1} \d J_1 + \xi_{1,2} \d J_2 \rangle|_{y_1(t)=x_1, y_2(t)=x_2} \\
            &= \lim_{\d t \to 0} \frac{1}{\d t} \left[ N_1 \d t + g_{1,1} \langle\d w_1\rangle + g_{1,2} \langle\d w_2 \rangle + \langle\xi_{1,1} \rangle \langle\d J_1 \rangle + \langle\xi_{1,2} \rangle \langle\d J_2 \rangle \right] \\
            &= N_1,
\end{align*}
where $\langle g_{i,j} \d W_j \rangle = \langle g_{i,j} \rangle \langle \d W_j \rangle = 0$, because a Wiener process has the property $\langle \d W_j \rangle =0$.
Further $\langle \xi_{i,j} \d J_j \rangle = \langle \xi_{i,j} \rangle \langle \d J_j \rangle = 0$, since $\xi_{i,j}$ is a Gaussian with zero mean, i.e., $\langle \xi_{i,j} \rangle = 0$.

Mutatis mutandis for $\Mtwo{0}{1}$.

\subsubsection{Kramers--Moyal coefficient $\Mtwo{1}{1}$}
\begin{align*}
\Mtwo{1}{1} &=  \lim_{\d t \to 0} \frac{1}{\d t} \langle (\d y_1)^1 (\d y_2)^1\rangle|_{y_1(t)=x_1, y_2(t)=x_2} \\
            &= \lim_{\d t \to 0} \frac{1}{\d t} \langle ( N_1 \d t + g_{1,1} \d w_1 + g_{1,2} \d w_2 + \xi_{1,1} \d J_1 + \xi_{1,2} \d J_2 ) \cdot \\
            & \quad \quad \quad ( N_2 \d t + g_{2,1} \d w_1 + g_{2,2} \d w_2 + \xi_{2,1} \d J_1 + \xi_{2,2} \d J_2 ) \rangle|_{y_1(t)=x_1, y_2(t)=x_2} \\
            &= \lim_{\d t \to 0} \left[N_1 N_2 \d t + g_{1,1} g_{2,1} \langle (\d w_1)^2 \rangle \frac{1}{\d t} + g_{1,2} g_{2,2} \langle (\d w_2)^2 \rangle \frac{1}{ dt} + \mathcal{O}(\dt) \right] \\
            &= g_{1,1} g_{2,1} + g_{1,2} g_{2,2},
\end{align*}
where higher-order terms $\mathcal{O}(\d t)^{\epsilon}$, with $\epsilon > 0$, vanish in the limit $\d t \to 0$.
Recall as well $\langle (\d w_i)^2 \rangle = \d t$.

\subsubsection{Kramers--Moyal coefficients $\Mtwo{2}{0}$ and $\Mtwo{0}{2}$}
\begin{align*}
\Mtwo{2}{0} &=   \lim_{\d t \to 0} \frac{1}{\d t}  \langle (\d y_1)^2 \rangle|_{y_1(t)=x_1, y_2(t)=x_2} \\
            &=   \lim_{\d t \to 0} \frac{1}{\d t} \langle (N_1 \d t + g_{1,1} \d w_1 + g_{1,2} \d w_2 + \xi_{1,1} \d J_1 + \xi_{1,2} \d J_2)^2 \rangle|_{y_1(t)=x_1, y_2(t)=x_2} \\
            &=  \lim_{\d t \to 0} \left[N_1^2 \d t + g_{1,1}^2 \langle (\d w_1)^2 \rangle \frac{1}{\d t} + g_{1,2}^2 \langle (\d w_2)^2 \rangle \frac{1}{\d t} + \langle \xi_{1,1}^2 \rangle \langle (\d J_1)^2 \rangle \frac{1}{\d t} + \right. \\
            & \left.\quad \quad \quad \langle \xi_{1,2}^2 \rangle \langle (\d J_2)^2 \rangle \frac{1}{\d t} + \mathcal{O}(\d t) \right]\\
            &= \left[g_{1,1}^2  + s_{1,1} \lambda_1  + g_{1,2}^2 + s_{1,2} \lambda_2\right],
\end{align*}
using the previously employed nomenclature $\langle \xi_{ij}^2 \rangle = \sigma_{\xi_{ij}}^2 = s_{ij}$ as well as $\langle (\d J_i)^2 \rangle = \lambda_i \d t$.

Mutatis mutandis for $\Mtwo{0}{2}$
\begin{align*}
\Mtwo{0}{2} =  \left[g_{2,1}^2 + s_{2,1} \lambda_1 + g_{2,2}^2  + s_{2,2} \lambda_2\right],
\end{align*}

\subsubsection{Kramers--Moyal coefficient $\Mtwo{2}{2}$}
\begin{align*}
\Mtwo{2}{2} &= \lim_{\d t \to 0} \frac{1}{\d t} \langle (\d y_1)^2 (\d y_2)^2 \rangle|_{y_1(t)=x_1, y_2(t)=x_2} \\
            &=\lim_{\d t \to 0}\frac{1}{\d t} \langle ( N_1 \d t + g_{1,1} \d w_1 + g_{1,2} \d w_2 + \xi_{1,1} \d J_1 + \xi_{1,2} \d J_2 )^2 \cdot \\
            & \quad \quad \quad ( N_2 \d t + g_{2,1} \d w_1 + g_{2,2} \d w_2 + \xi_{2,1} \d J_1 + \xi_{2,2} \d J_2 )^2 \rangle|_{y_1(t)=x_1, y_2(t)=x_2} \\
            &= \lim_{\d t \to 0}\frac{1}{\d t} \bigg[ \mathrm{terms}(N_1, N_2, \mathcal{O}(\d t^4)) + \mathrm{terms}(g_{ij}, \mathcal{O}(\d t^2)) + \mathrm{terms}(\mathrm{mixing \ } \xi_{ij}) + \\
            & \quad \quad \quad \langle \xi_{1,1}^2 \rangle \langle \xi_{2,1}^2 \rangle  \langle (\d J_1)^4 \rangle + \langle \xi_{1,2}^2 \rangle \langle \xi_{2,2}^2 \rangle \langle (\d J_2)^4 \rangle + \\
            & \quad \quad \quad \langle \xi_{1,1}^2 \rangle \langle \xi_{2,2}^2 \rangle \langle (\d J_1)^2 \rangle \langle (\d J_2)^2 \rangle + \langle \xi_{1,2}^2 \rangle \langle \xi_{2,1}^2 \rangle \langle (\d J_1)^2 \rangle \langle (\d J_2)^2 \rangle \bigg]\\
            &=  \left[s_{1,1} s_{2,1} \lambda_1 + s_{1,2} s_{2,2} \lambda_2\right].
\end{align*}
Terms including $\d t$ on the right-hand side of the above equation vanish for $\d t \to 0$, where as well $\langle \xi_{1,1} \xi_{1,2} \rangle = \langle \xi_{1,1} \rangle \langle \xi_{1,2} \rangle = 0$, and $\frac{1}{\d t}[\langle (\d J_1)^2 \rangle \langle (\d J_2)^2 \rangle] = \frac{1}{\d t}[\lambda_1 \d t \lambda_2 \d t] \propto \d t$ vanishes in the limit $\d t \to 0$.

\subsubsection{Kramers--Moyal coefficients $\Mtwo{\ell}{m}$, for $2\times(\ell, m) \geq 2$}
For $(2\ell, 2m)$, with $(\ell, m)\geq 4$, the Kramers--Moyal coefficients $\Mtwo{2\ell}{2m}$ are as follows
\begin{align*}
\Mtwo{2\ell}{2m} &=\lim_{\d t \to 0}  \frac{1}{\d t} \langle (\d y_1)^{2\ell} (\d y_2)^{2m} \rangle|_{y_1(t)=x_1, y_2(t)=x_2} \\
            &= \lim_{\d t \to 0} \frac{1}{\d t}  \langle ( N_1 \d t + g_{1,1} \d w_1 + g_{1,2} \d w_2 + \xi_{1,1} \d J_1 + \xi_{1,2} \d J_2 )^{2\ell} \cdot \\
            & \quad \quad \quad ( N_2 \d t + g_{2,1} \d w_1 + g_{2,2} \d w_2 + \xi_{2,1} \d J_1 + \xi_{2,2} \d J_2 )^{2m} \rangle|_{y_1(t)=x_1, y_2(t)=x_2} \\
            &= \lim_{\d t \to 0} \frac{1}{\d t} \left[ \langle \xi_{1,1}^{2\ell} \rangle \langle \xi_{2,1}^{2m} \rangle \langle (\d J_1)^{2(\ell+ m)} \rangle + \langle \xi_{1,2}^{2\ell} \rangle \langle \xi_{2,2}^{2m} \rangle \langle (\d J_2)^{2(\ell+ m)} \rangle \right] \\
            &=\left[\langle \xi_{1,1}^{2\ell} \rangle \langle \xi_{2,1}^{2m} \rangle \ \lambda_1 + \langle \xi_{1,2}^{2\ell} \rangle \langle \xi_{2,2}^{2m} \rangle \ \lambda_2\right] \\
            &= \frac{(2\ell)!}{2^\ell \ell!} \frac{(2m)!}{2^m m!}\left[ s_{1,1}^\ell s_{2,1}^m \lambda_1 + s_{1,2}^\ell s_{2,2}^m \lambda_2 \right].
\end{align*}
In the last step, take the fact that the jump amplitudes $\xi_{i,j}$ are Gaussian distributed, thus $\langle \xi_{i,j}^{2\ell} \rangle \propto \sigma_{\xi_{i,j}}^{2\ell} = s_{i,j}^{\ell}$.
In this manner, all Kramers--Moyal coefficients $\Mtwo{2\ell}{2m}$, with $(\ell, m) \geq 1$ are obtained.

\section{Extended results for modeled data by Eq.~(\ref{eq:Cubib_JD_ms})}\label{App:Cubib_JD_ms}
\begin{figure*}[h!]
\includegraphics[width=.32\linewidth]{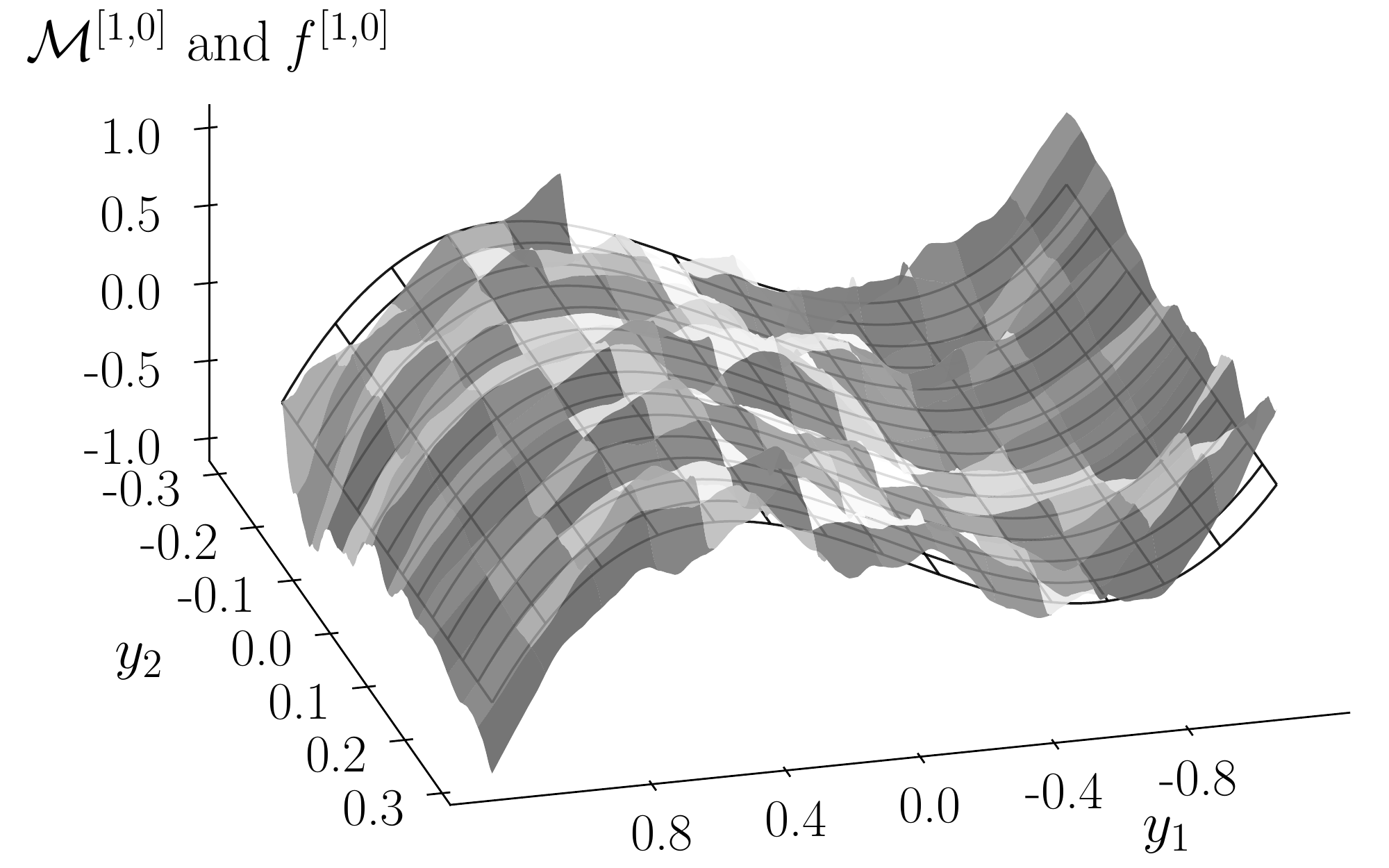}
\includegraphics[width=.32\linewidth]{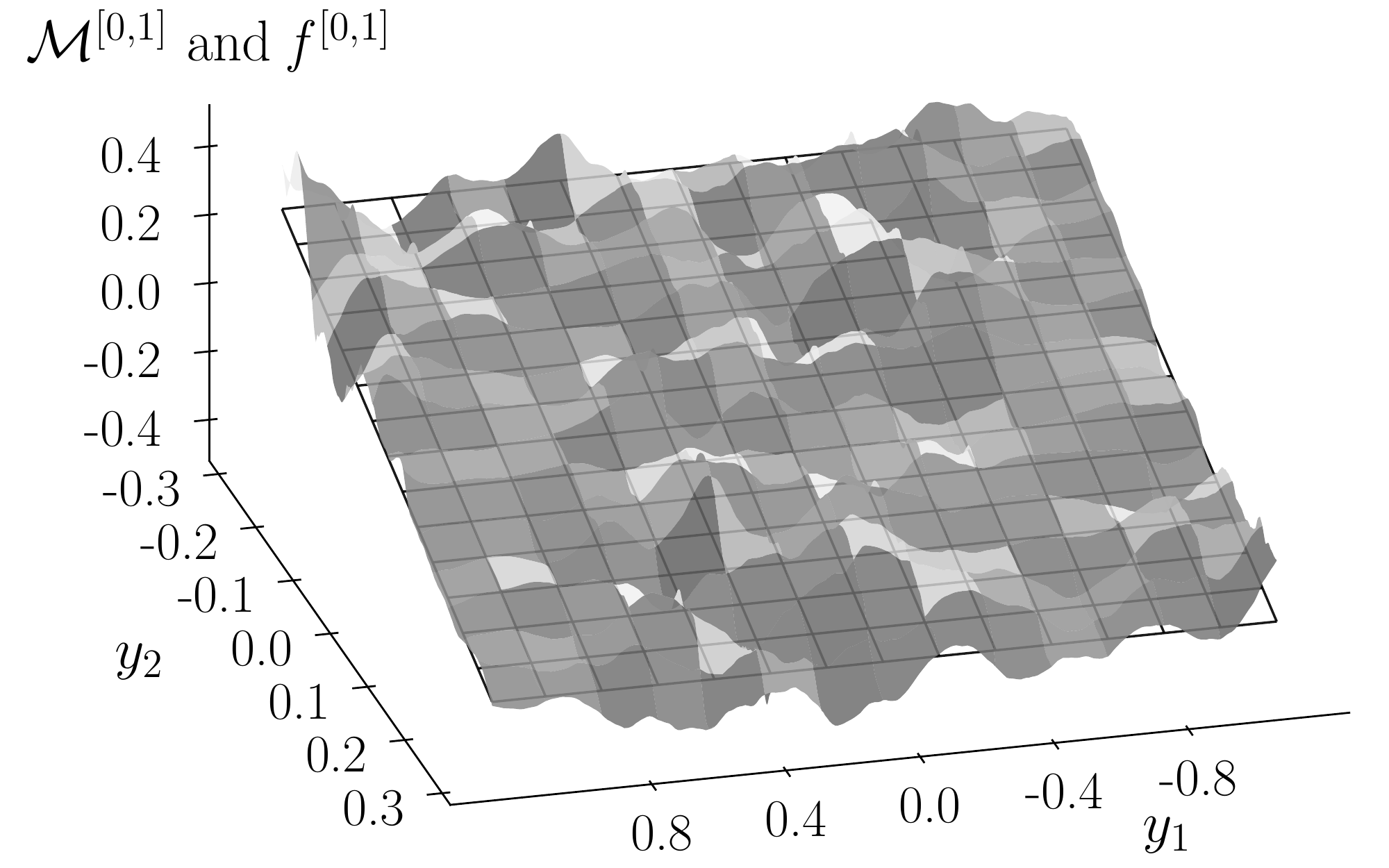}
\includegraphics[width=.32\linewidth]{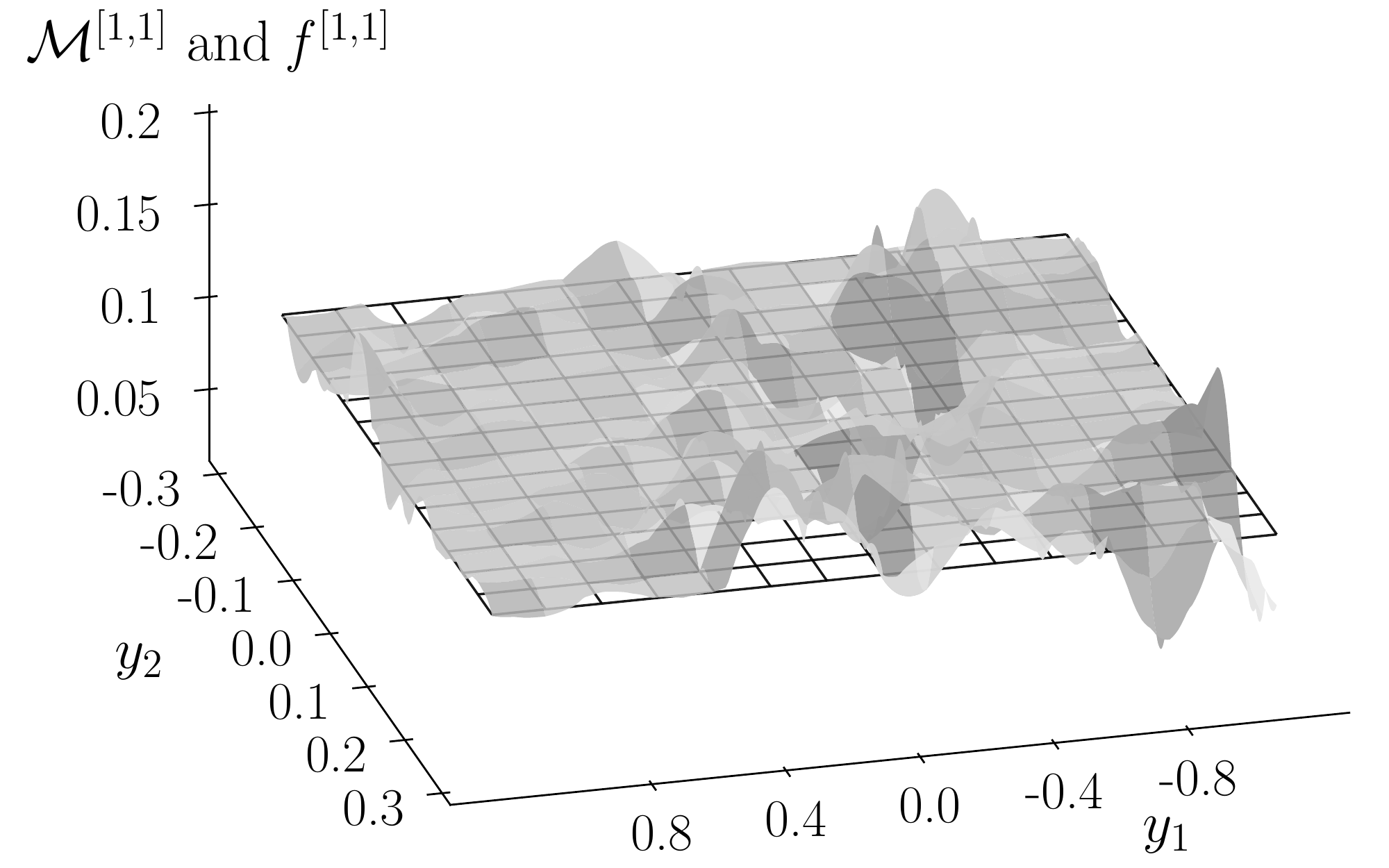}
\includegraphics[width=.32\linewidth]{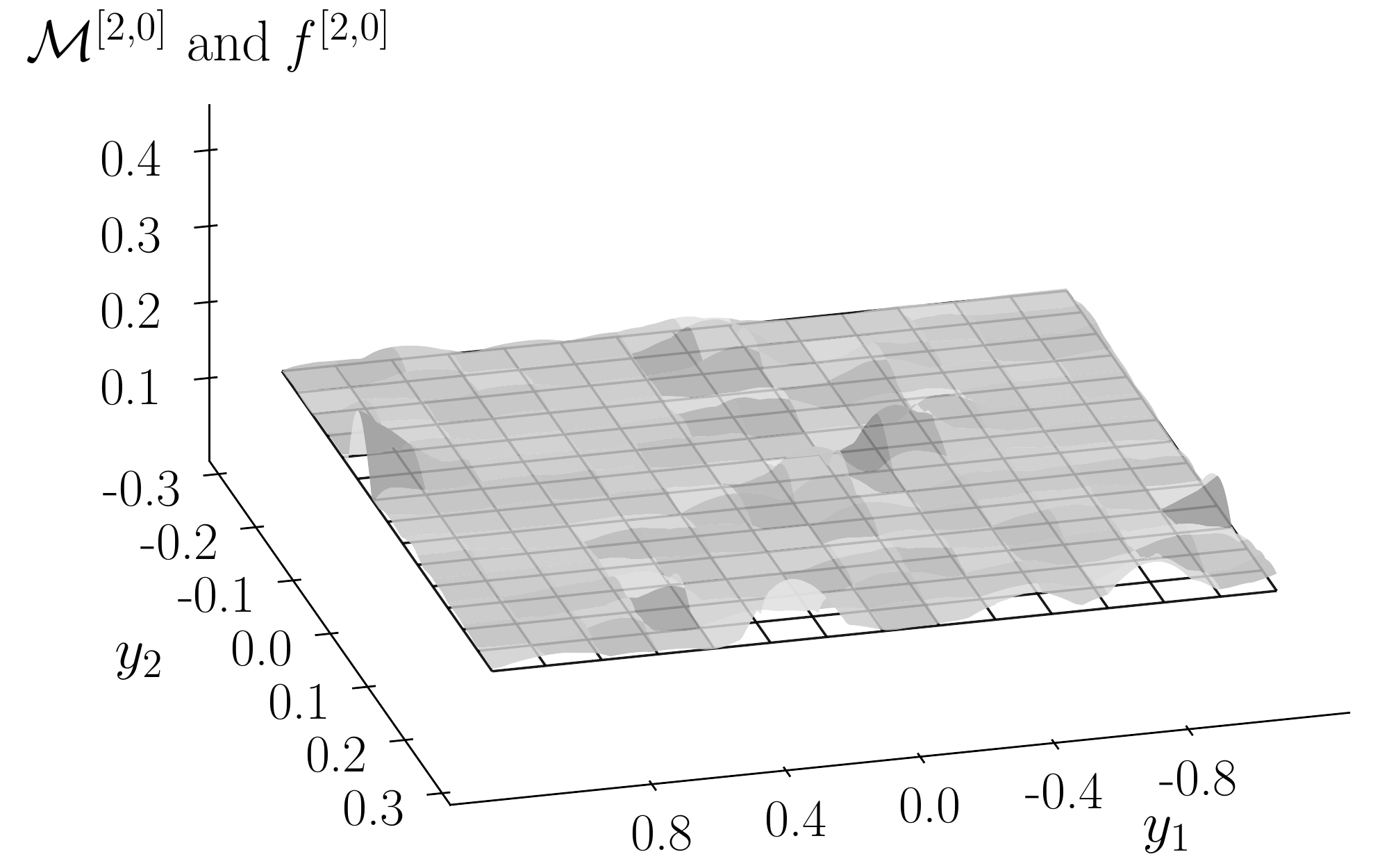}
\includegraphics[width=.32\linewidth]{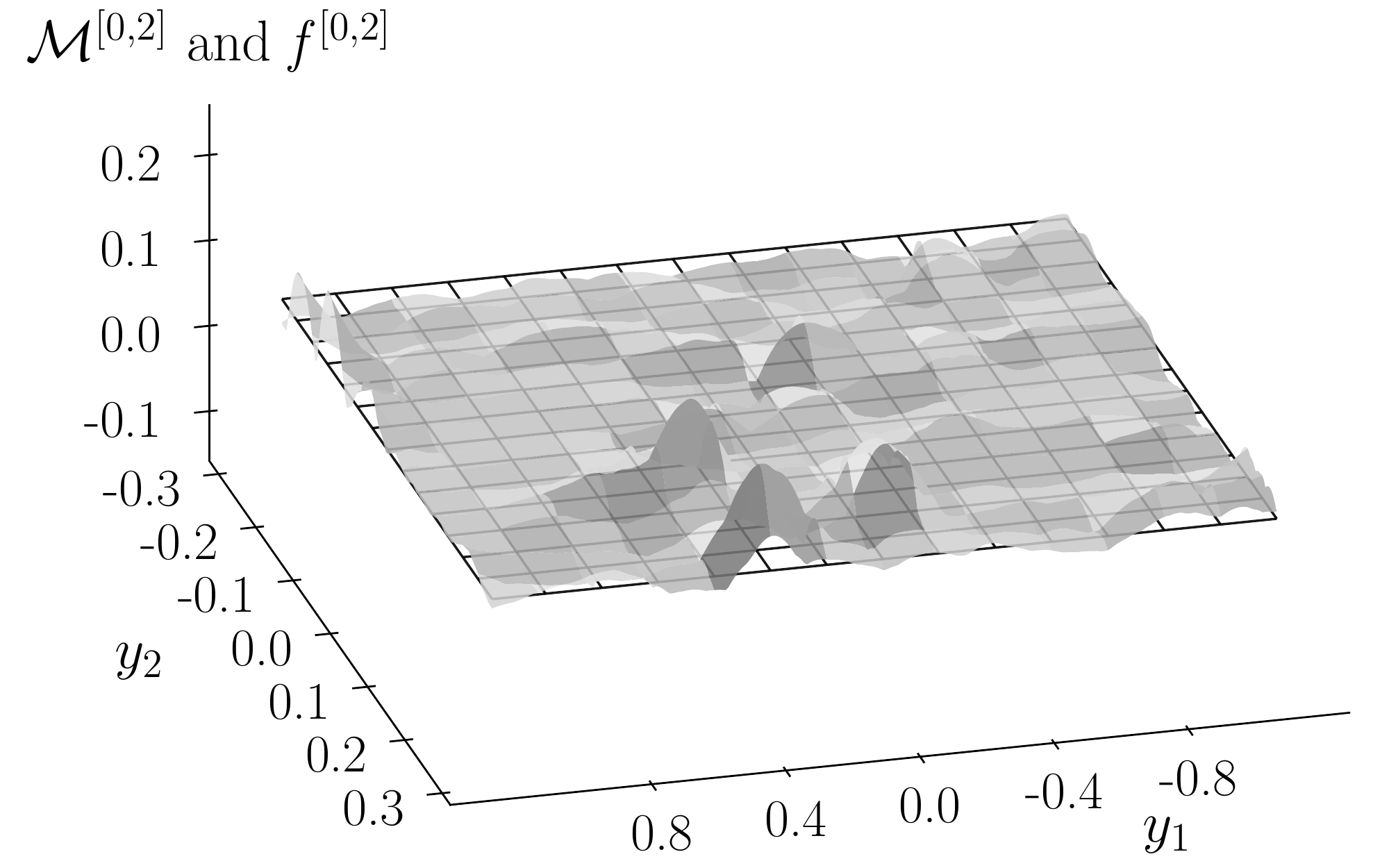}
\includegraphics[width=.32\linewidth]{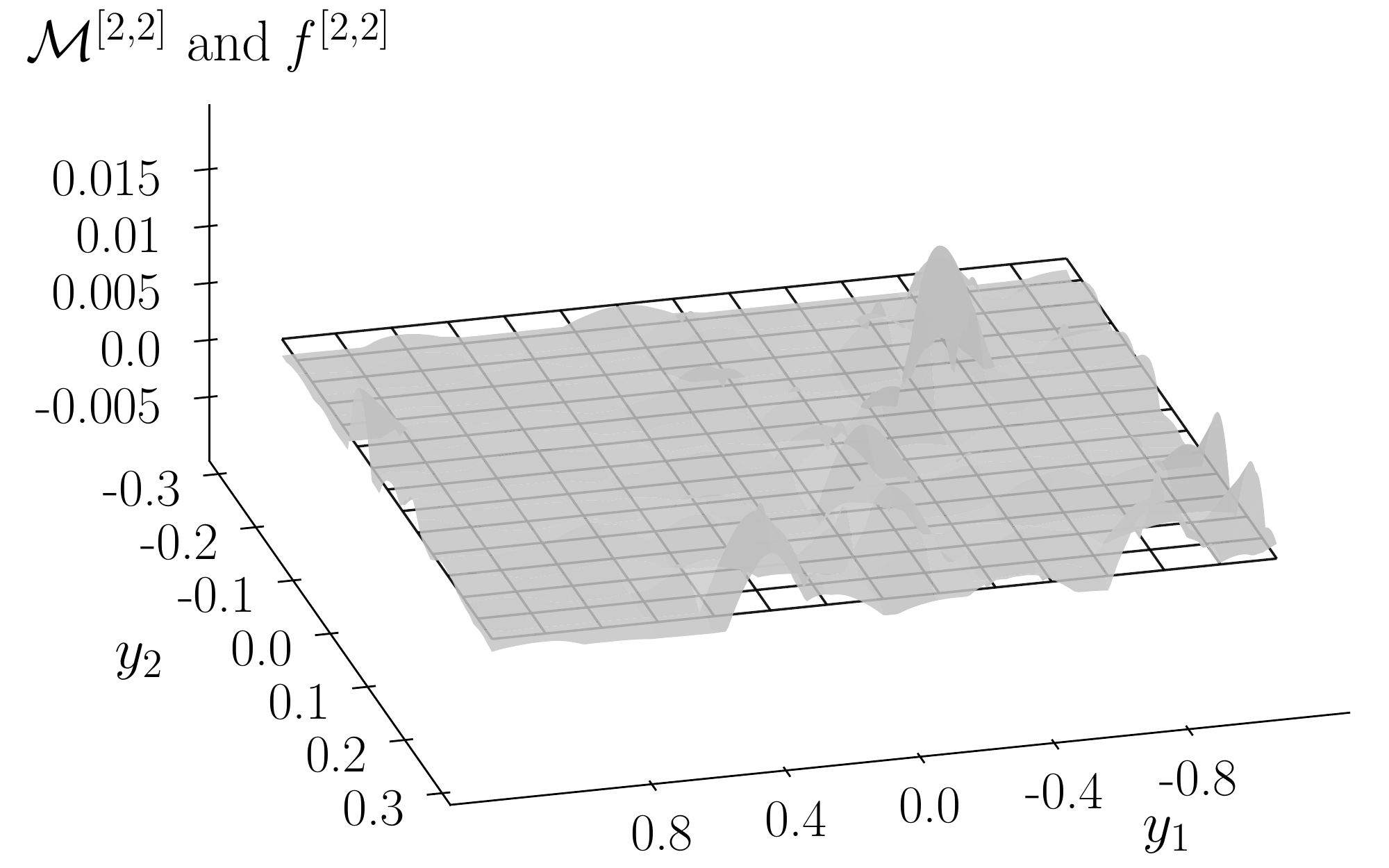}
\includegraphics[width=.32\linewidth]{Cubic_drift_JD_ms_40_and_f}
\includegraphics[width=.32\linewidth]{Cubic_drift_JD_ms_04_and_f}
\includegraphics[width=.32\linewidth]{Cubic_drift_JD_ms_44_and_f}
\includegraphics[width=.32\linewidth]{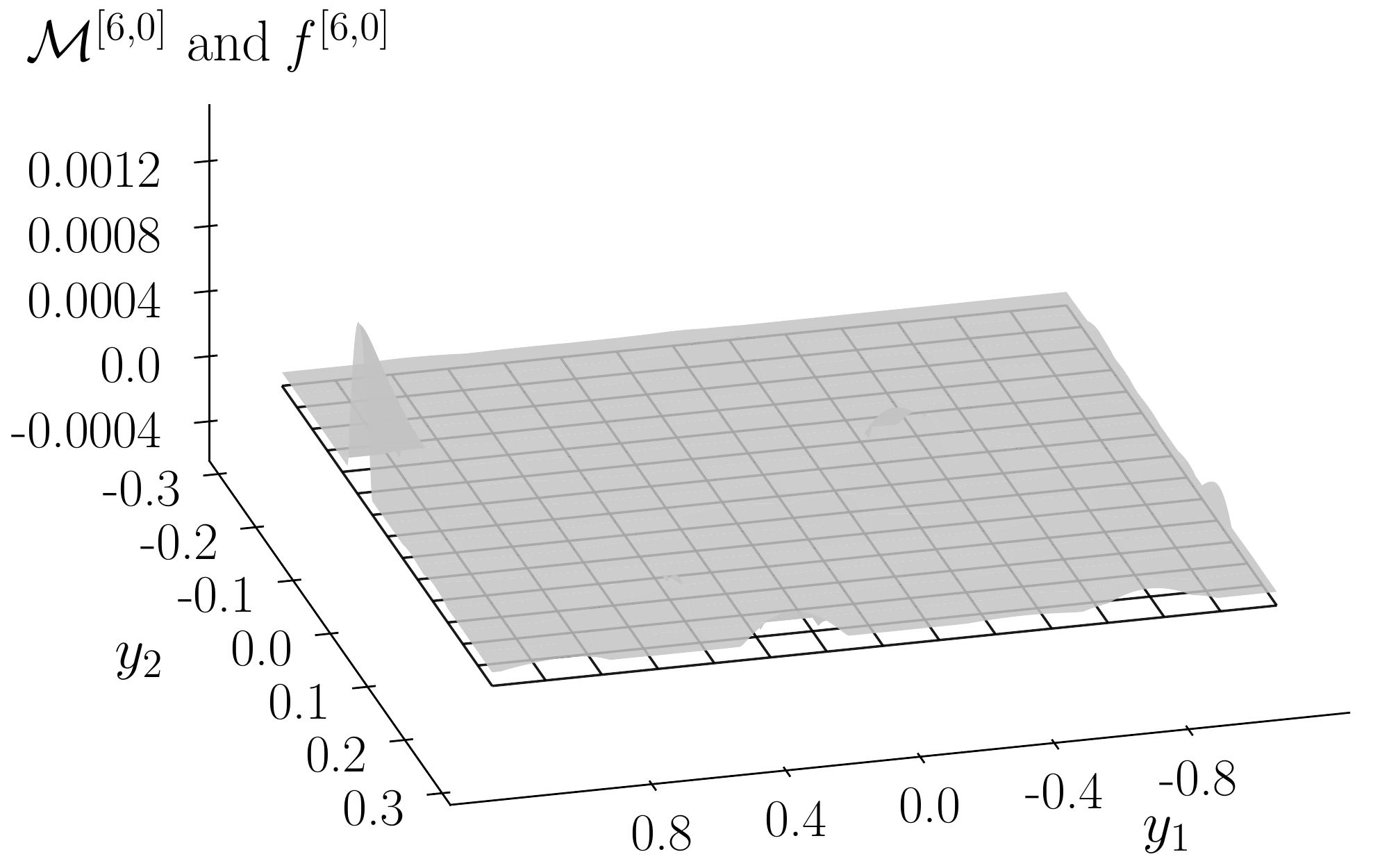}
\includegraphics[width=.32\linewidth]{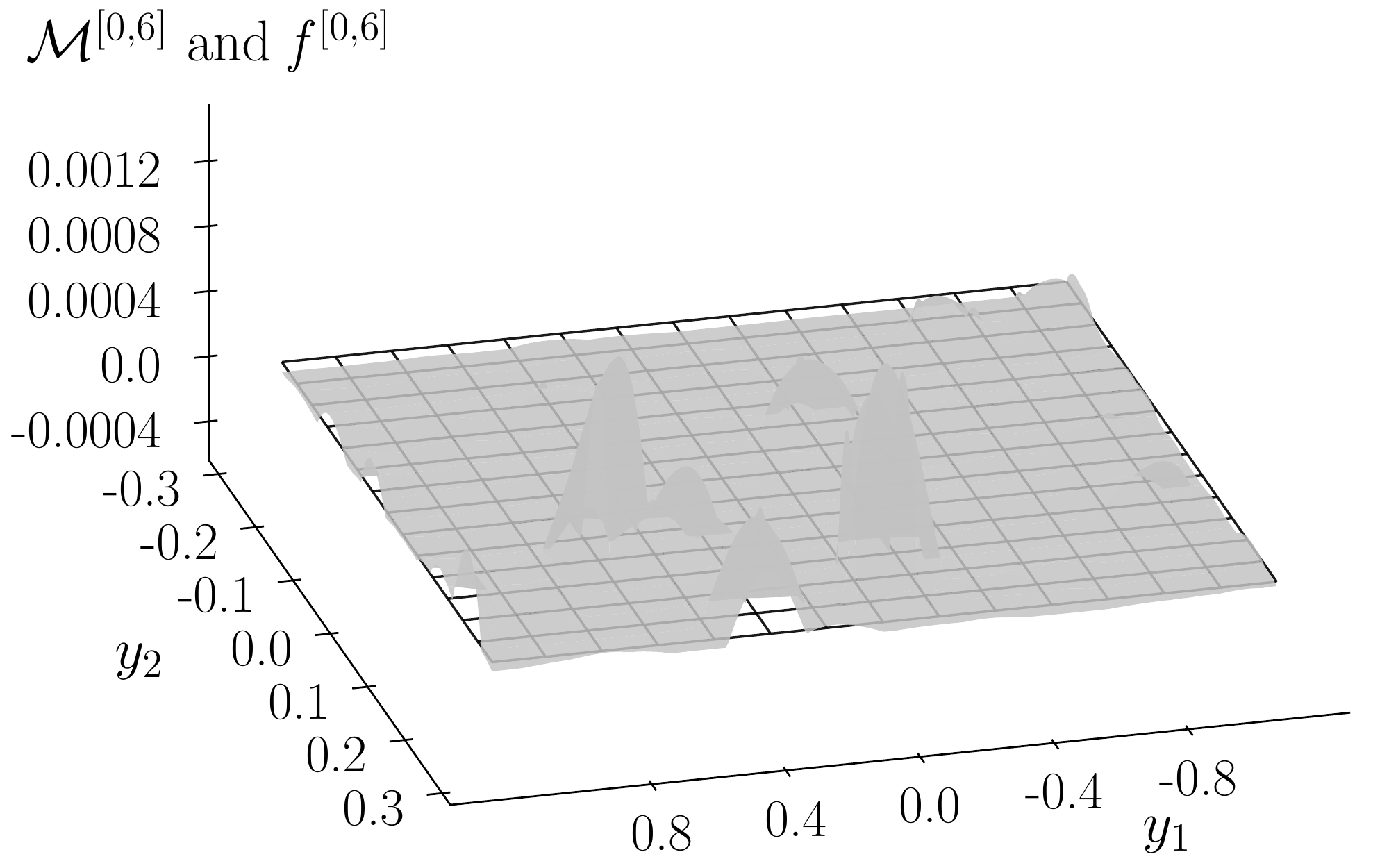}
\includegraphics[width=.32\linewidth]{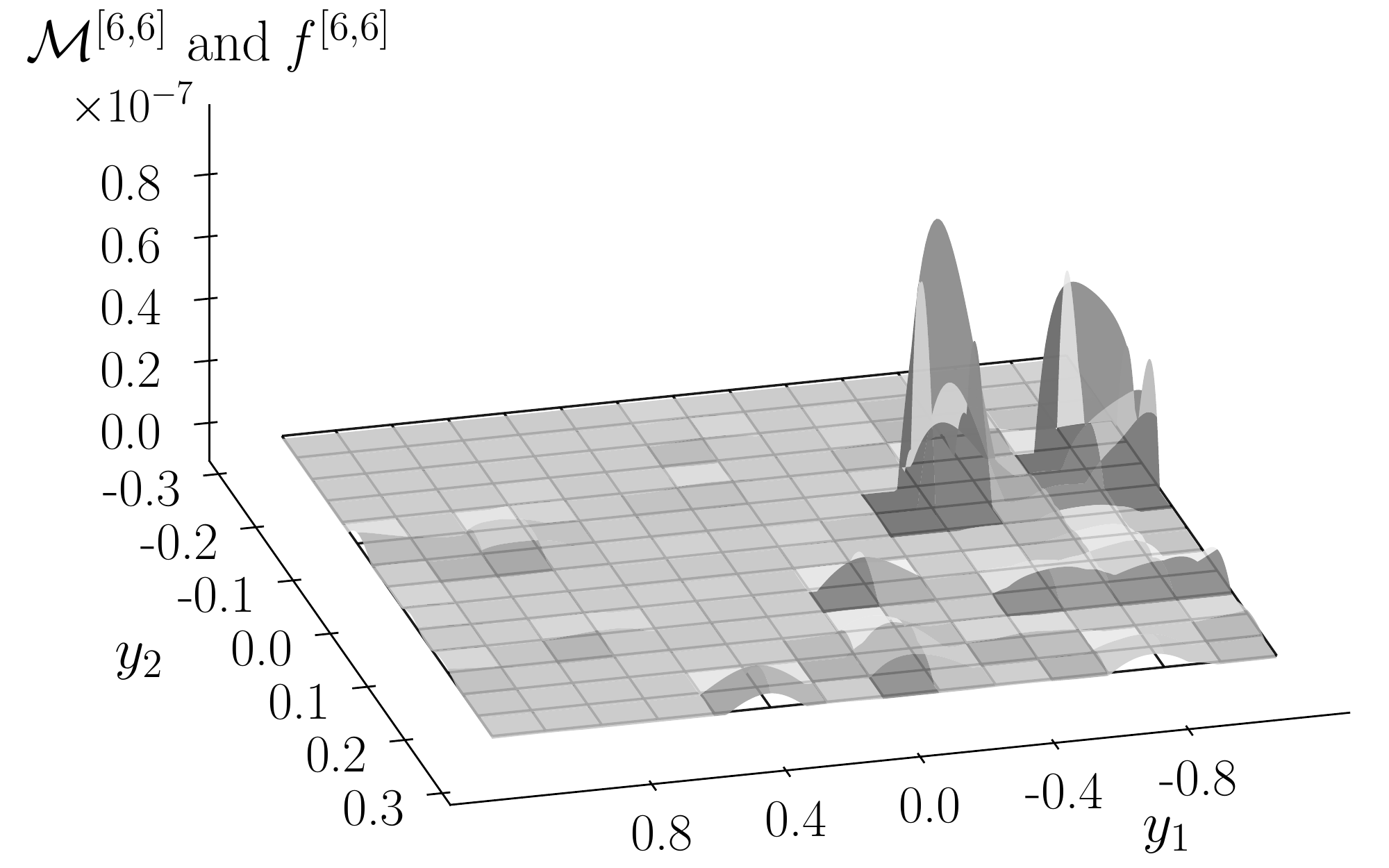}
\includegraphics[width=.32\linewidth]{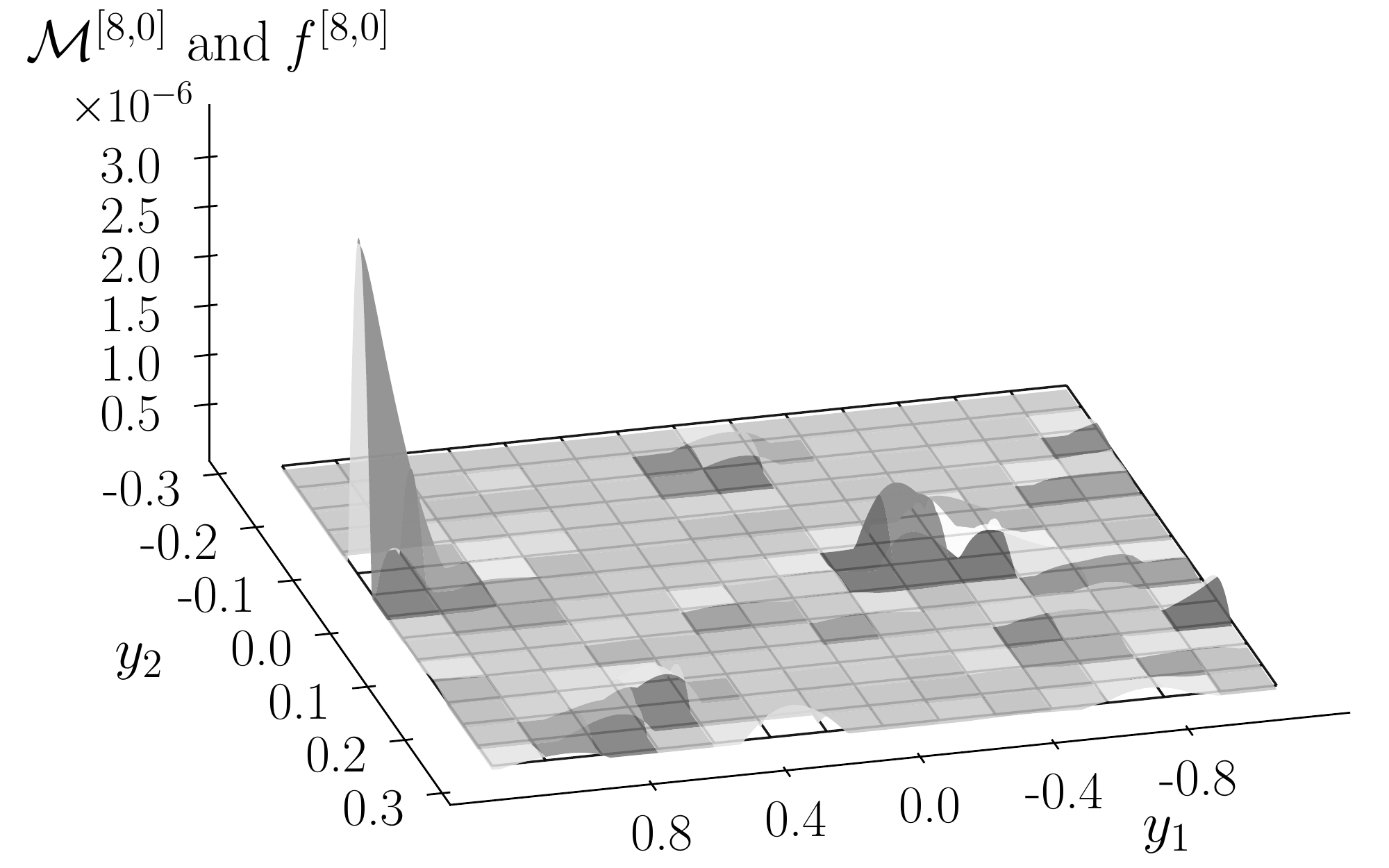}
\includegraphics[width=.32\linewidth]{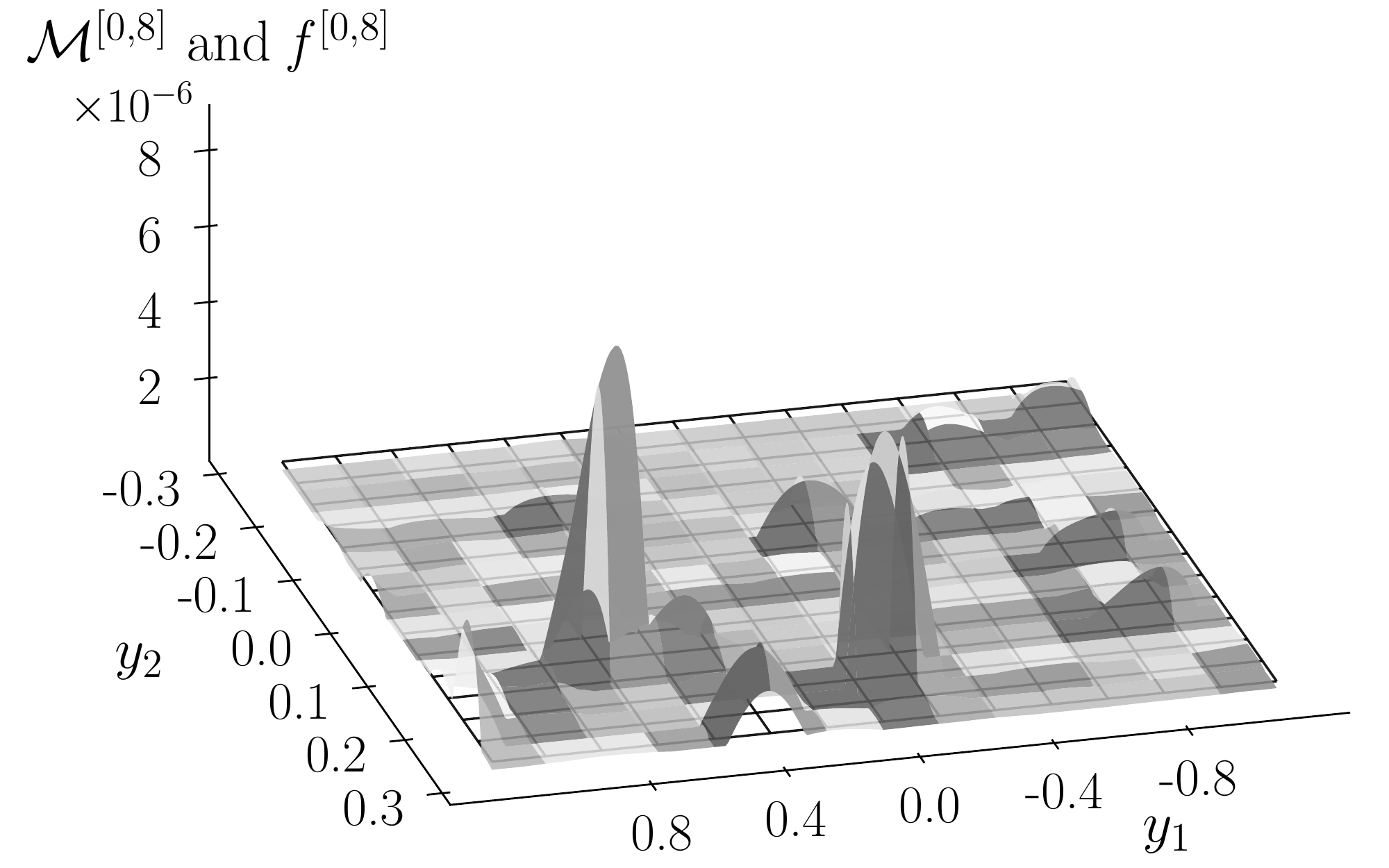}
\caption{Two-dimensional Kramers--Moyal (KM) coefficients $\Mtwo{\ell}{m}$ of bivariate jump-diffusion processes given by Eq.~\eqref{eq:Cubib_JD_ms} and all theoretical expected functions $\ftwo{\ell}{m}$ associated with each KM coefficient according to Eqs.~\eqref{eq:drift_km2}, \eqref{eq:diffusion_km2}, and \eqref{eq:jumps_km2}.
Shown are the KM coefficients $\Mtwo{1}{0}$, $\Mtwo{0}{1}$, $\Mtwo{1}{1}$, $\Mtwo{2}{0}$, $\Mtwo{0}{2}$, $\Mtwo{2}{2}$, $\Mtwo{4}{0}$, $\Mtwo{0}{4}$, $\Mtwo{4}{4}$, $\Mtwo{6}{0}$, $\Mtwo{0}{6}$, $\Mtwo{6}{6}$, $\Mtwo{8}{0}$, and $\Mtwo{0}{8}$.
The respective error volumes read: $V_\text{err}^{[1,0]}= 0.6836$, $V_\text{err}^{[0,1]}= 0.23$, $V_\text{err}^{[1,1]}=0.01$, $V_\text{err}^{[2,0]}= 0.01$, $V_\text{err}^{[0,2]}=0.03$, $V_\text{err}^{[2,2]}= 0.01$, $V_\text{err}^{[4,0]}=0.02$, $V_\text{err}^{[0,4]}= 0.03$,  $V_\text{err}^{[4,4]}<0.01$, $V_\text{err}^{[6,0]}= 0.02$, $V_\text{err}^{[0,6]}=0.01$, $V_\text{err}^{[6,6]}= 0.01$, $V_\text{err}^{[8,0]}=0.04$, $V_\text{err}^{[0,8]}= 0.26$.}
\label{fig:Cubic_JD_ms}
\end{figure*}

\end{appendices}

\end{document}